\documentclass[12pt,pre,aps,superscriptaddress]{revtex4}
\usepackage{graphicx}
\newcommand{\beq}[2]{\begin{equation}#1\label{#2}\end{equation}}
\newcommand{\ceq}[1]{(\ref{#1})}

\newfont{\mbld}{cmbx10 scaled 800}
\newfont{\cab}{cmsy10 scaled 1200}
\newfont{\scab}{cmsy10 scaled 1000}
\newfont{\bcall}{cmbsy10 scaled 1200}

\begin{document}
\title{A path integral approach to the dynamics of a random chain
with rigid constraints}
\author{Franco Ferrari}
\email{ferrari@univ.szczecin.pl}
\author{Jaros{\l}aw Paturej}\email{jpaturej@univ.szczecin.pl}
\affiliation{Institute of Physics and CASA*, University of Szczecin,
  ul. Wielkopolska 15, 70-451 Szczecin, Poland}
%\author{Vakhtang G. Rostiashvili}\email{rostiash@mpip-mainz.mpg.de}
\author{Thomas A. Vilgis} 
\email{vilgis@mpip-mainz.mpg.de}
\affiliation{Max Planck Institute for Polymer Research, 10
  Ackermannweg, 55128 Mainz, Germany}

\begin{abstract}
In this work the dynamics of a chain consisting of a set of
beads attached to the ends of segments of fixed lengths is
investigated.
The chain fluctuates at constant temperature in a viscous medium.
For simplicity, all interactions among the beads have been switched
off and the number of spatial dimensions has been limited to two. 

In the limit in which the chain becomes a continuous system, its
behavior may be described by a path integral, in which the rigid
constraints coming from the infinitesimally small segments are imposed
by means of a functional delta--function. In this way a model of the
dynamics of the chain is obtained, which closely resembles a
two-dimensional nonlinear sigma model. The partition function of this
generalized nonlinear sigma model is computed explicitly 
for a ring-shaped chain in the
semiclassical approximation.
The behavior of the chain at both long and short scales of time and distances
is investigated. The connection
  between the generalized nonlinear sigma model presented here and the
  Rouse model is  discussed. 
\end{abstract}
\maketitle
\section{Introduction}\label{sec:intro}
Subject of this work is a study of the dynamics of a continuous chain
which is subjected to thermodynamic fluctuations at constant
temperature $T$. The chain is represented as the limit of a system of
beads and links of fixed length, in which the number $N$ of beads
becomes infinite, while the length $a$ of the links goes to zero in
such a way that the total length of the chain $L=Na$ remains a finite
constant. 
Problems of this kind are encountered for example in polymer physics,
because with some approximation long flexible polymers may be regarded
as continuous chains \cite{doiedwards}. 
It is thus spontaneous to
consider an isolated 
polymer fluctuating in some viscous environment
 as a concrete realization of the system
investigated here. 
A few
applications in which the dynamics of a chain turns out to be
relevant are mentioned in Ref.~\cite{EdwGoo}. 
In particular, in Ref.~\cite{EdwGoo} six regimes of the chain dynamics
are distinguished, which apply not only
to the well known case of a polymer in a solution,
but also to other cases, like for instance
those of an isolated cold chain and
of a hot polymer 
in the vapor phase.

From polymer physics, one may borrow the standard approach to the
dynamics of a chain. It consists in considering the
fluctuations of the chain as a stochastic process, which is usually
described with the help of the Langevin equations or, alternatively,
of the Fokker--Planck--Smoluchowski equations \cite{doiedwards}.
This approach
leads to the well known models of Rouse \cite{rouse} and Zimm \cite{zimm}
which allow a satisfactory understanding of the main properties of polymers in
solutions. One major drawback of these coarse grained models is that they
do not take into account the rigid constraints which are necessary
 in order to keep constant the length of the
links connecting the beads. The Rouse and Rouse--Zimm 
equation consider 
instead beads joined together by springs, where the local spring is
infinitely extensible. In this way the length of the chain
is not fixed and is allowed to become infinite. Moreover, in the
continuous limit the Rouse equation is nothing but the stochastic
equation (Langevin equation) for the
classical Wiener measure, which yields paths
without  well defined tangent vectors
\cite{doiedwards}. These problems have
been tackled by various attempts, see e.g. \cite{yamakawa,stockmayer}.
However, the correct use of rigid constraints in (stochastic) dynamics requires
some mathematical effort \cite{arti, zinn}, in contrast to the static cases
where rigid constraints can be implemented by Dirac delta functions in the
partition functions. For instance, the
probability distribution of $N$ ideal closed chains topologically
linked together may be represented as a path integral of $N$
noninteracting particles with the insertion of Dirac delta functions
which enforce the topological constraints on their trajectories
\cite{edwa,vilbre, FeKlLa, kleinertpi}.

Here a strategy similar to that used for static chains will be applied
to dynamics. We consider
the distribution $\Psi_{disc}$ of the
probability that a fluctuating chain passes from a initial discrete spatial
conformation $\Gamma_i$ to a final one $\Gamma_f$ in a given interval
of time $\Delta t=t_f-t_i$. 
The idea behind our approach is based on the fact that 
the beads of the chain may be regarded as a set of $N$ Brownian
particles with constrained trajectories.
The constraints arise due to the presence of the $N-1$ links of
fixed lengths connecting the beads. As a consequence, it is possible
to write the probability distribution
$\Psi_{disc}$ in the form of a path integral describing the
fluctuations of $N$  Brownian particles with the insertion of
  Dirac delta functions. The latter are needed in order to impose
 the necessary conditions on the
  trajectories of the particles.
For simplicity,
possible interactions among the beads have been switched off, including
hydrodynamic interactions and only the two-dimensional case has been
discussed. 

The limit in which the chain 
becomes continuous is not entirely trivial.
It involves the
vanishing of three crucial quantities, the mass of the beads, their
mobility and their size. 
After performing this limit very carefully, we obtain as a final
result the probability
distribution $\Psi$ of the continuous chain. It is found that $\Psi$
 consists of a path integral which closely
resembles the partition function of a two-dimensional nonlinear sigma
model~\cite{nlsigma}. 
For this reason the
obtained model will be called the generalized nonlinear 
sigma model (GNLSM).
The main difference 
from the 
nonlinear sigma model
is that the holonomic
constraint  is replaced in the GNLSM by an
nonholonomic constraint, which requires that the tangent at every point
of the trajectory of the chain is an unit vector. 
 The Lagrange multiplier that imposes this
nonholonomic  constraint plays the same role of pressure in the
hydrodynamics of 
incompressible fluids, a fact already noted in Ref.~\cite{EdwGoo}.
The ``incompressibility'' is related to the
fact that it is not possible to ``compress'' the lengths of the links
joining the beads.
The GNLSM is well suited to study all situations in which
it is possible to 
assume that the thermodynamic 
fluctuations
and the conformational changes are small and slow.
Systems satisfying these requirements are for instance cold
chains or chains in a very viscous medium.

In principle, the GNLSM is exactly solvable after performing two
gaussian integrations, but the presence
of non-trivial boundary conditions complicates the calculation of
the
probability distribution $\Psi$. Even the
method of the effective potential, which is useful
to investigate phase transitions in nonlinear sigma models,
cannot be easily applied.
As a matter of fact, in the nonlinear sigma model
the effective potential 
is computed assuming that
the field configurations which minimize the action are constants. However,
in the GNLSM configurations of this kind
correspond to the situation in which the chain has collapsed to a
point and are thus unphysical.
Despite these difficulties, it is possible to compute
the probability distribution $\Psi$ and the
associated generating functional of the correlation functions of the
bond vectors $\Psi[J]$ using a background field approximation, in which
small gaussian fluctuations are considered in the background of a
dominating 
 classical conformation. The initial and final conformations of the
 chain are 
picked up by choosing in a suitable way the background classical
solution and by tuning the boundary conditions of the gaussian
fluctuations. 
The semiclassical approximation is valid in the case of low
temperatures or of 
highly viscous media,  exactly the regimes in which the GNLSM
can be applied.
The explicit formulas of $\Psi$
and $\Psi[J]$ derived here show
that the 
fluctuations which deform the chain 
along directions which are tangent to the trajectory of the classical
background conformation propagate differently from the normal
fluctuations. 
This fact is a direct consequence of the presence of rigid
constraints and could be relevant for example in the theory of
formation of single polymer crystals \cite{lnp}. 

The material presented in this paper is organized as follows.
In Section II the form of the Lagrangian of a classical discrete chain
in two dimensions has been derived in polar and in cartesian
coordinates. 
No restrictions are posed to the motion of the
discrete chain. The limit to a continuous chain is however performed assuming
that one of the ends of the chain is fixed.
The calculation in polar coordinates shows that one crucial term in
the Lagrangian disappears in the continuous limit. This fact simplifies
the classical equations of motion of the chain. 
The probability distribution $\Psi$ of the fluctuating chain
is computed in Section III using a path
integral approach. Some subtleties arising when taking  the continuous
limit in
the probability distribution of the discrete chain are discussed.
Section IV is dedicated to the study of the classical solutions of the
GNLSM. 
It is shown that
the only possible classical 
solutions are  time independent, apart from rigid
translations with constant velocity of the whole chain.
The computation of the probability distribution 
$\Psi$ and of the generating functional $\Psi[J]$
for a ring-shaped chain
is performed 
in the semiclassical approximation in Section V.
The physical interpretation of the results obtained in the previous
Sections is presented in Section VI.
The equilibrium limit of the GNLSM and its connection with the Rouse
model
are studied in Section VII.
Finally, in Section VIII the conclusions are drawn and  possible
future developments are discussed.

\section{The classical dynamics of a continuous chain}\label{sec2}

\begin{figure}[bpht]
\centering
\includegraphics[width=.5\textwidth]{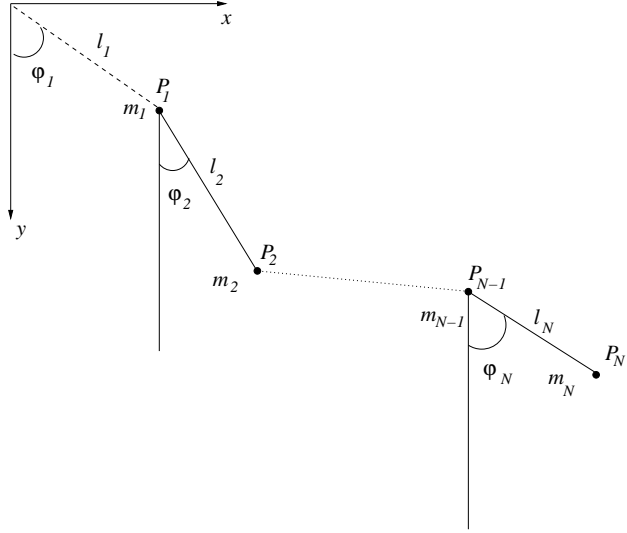}
\caption{\footnotesize
A chain with $N$ segments. Let us note that the end point
  $P_1$ is not bound to stay at a fixed distance with respect to the
  origin of the cartesian reference system.
}\label{system2y}
\end{figure}

Let us consider a discrete chain of $N-1$ segments of fixed lengths
$l_2,\ldots,l_N$ in 
the two-dimensional plane. Each segment $P_{i+1}P_i$ is completely specified
by the positions of its end points $P_{i+1}$ and $P_i$. In cartesian
coordinates $(x,y)$ these positions are given by the radius vectors:
\beq{\mathbf R_i=(x_i,y_i)\qquad\qquad i=1,\ldots,N}{pointcoor} The segments
are joined together at the points $P_i$, where $2\le i\le N-1$, see
Fig.~\ref{system2y}, while $P_1$ and $P_N$ are the ends of the chain.
Moreover, at each point $P_i$, with $i=1,\ldots,N$, a mass $m_i$ is attached.
In the following we restrict ourselves to the case of a free chain. We will
see below that the addition of interactions is straightforward.

At this point we compute the kinetic energy of the above system:
\beq{K_{disc}=\sum_{i=1}^N
\frac{m_i}{2}\left(\dot x^2_i+\dot
  y^2_i\right)
}{totenechadef} 
The subscript $disc$ in the left hand side of Eq.~\ceq{totenechadef}
is to recall that we are considering at the moment a discrete chain
with $N-1$ segments.
For future purposes, it will be convenient to introduce the kinetic
energy of point $P_i$:
\beq{K_i=\frac{m_i}{2}\left(\dot x^2_i+\dot
  y^2_i\right)}{totenechadefdef}
Of course, to Eq.~\ceq{totenechadef} % and \ceq{totenechadefdef} 
 one should also add the constraints
\beq{
(x_i-x_{i-1})^2+(y_i-y_{i-1})^2=l_i^2\qquad\qquad i=2,\ldots,N
}{constrdiscln}
in order to enforce the requirement that the segments have fixed
length $l_i$. It is possible to eliminate these constraints
passing to 
polar coordinates: 
\beq{
  x_i=\sum_{j=1}^il_j\cos\varphi_j\qquad\qquad y_i=\sum_{k=1}^i
  l_j\sin\varphi_j\qquad\qquad i=1,\ldots,N }{carpol2} 
$\varphi_j$ is the
angle formed by segment $j$ with the $y-$axis, see
Fig.~\ref{system2y}.  According to our settings,
the radial coordinates $l_j$ are constants for $j=2,\ldots,N$. 
The length $l_1$, which denotes the distance of the end point
$x_1,y_1$ from the origin, is not fixed, so that $l_1=l_1(t)$ is
allowed to vary with the time $t$.  From Eq.~\ceq{carpol2} the velocity
components of the $i-$th segment may be written as follows:
\begin{eqnarray}
\dot
x_i&=&-\sum_{j=1}^{i-1}l_j\dot\varphi_j\sin\varphi_j-l_i
\dot\varphi_i\sin\varphi_i  
+\dot l_1\cos\varphi_1
\qquad\qquad i=2,\ldots,N\label{velcom1} \\
\dot
y_i&=&\sum_{j=1}^{i-1}l_j\dot\varphi_j\cos\varphi_j+l_i
\dot\varphi_i\cos\varphi_i
+\dot l_1\sin\varphi_1
\qquad\qquad i=2,\ldots,N \label{velcom2}\\
\dot x_1&=&-l_1\dot\varphi_1\sin\varphi_1+\dot
l_1\cos\varphi_1\label{velcom3}\\
\dot y_1&=&l_1\dot\varphi_1\cos\varphi_1+\dot
l_1\sin\varphi_1\label{velcom4}
\end{eqnarray}
Separating the contribution coming from the first $i-1$ variables as
shown in Eqs.~(\ref{velcom1}--\ref{velcom2}), the kinetic
energy $K_i$ of the $i-$th point can be expressed in terms of the kinetic
energy $K_{i-1}$ of the $(i-1)-$th segment: 
\beq{
  K_i=\frac{m_i}{m_{i-1}}K_{i-1}+\frac{m_i}{2}l_i^2\dot\varphi_i^2+
  m_i\sum_{j=1}^{i-1}l_il_j\dot\varphi_i\dot\varphi_j\cos(\varphi_j-\varphi_i)
  +m_il_i\dot\varphi_i\dot l_1\sin(\varphi_1-\varphi_i) }{kinenerecrel} It is
possible to solve the above recursion relation and to find a closed
expression of $K_i$ in polar coordinates. 
If we do that, at the end the total kinetic energy of the discrete
chain
becomes:
\begin{eqnarray}
K_{disc}&=&\frac M2\left(
l_1^2\dot\varphi_1^2+\dot l_1^2
\right )
+l_1\dot\varphi_1\sum_{i=1}^N
\sum_{j=1}^{i-1}m_i
l_{i-j+1}
\dot\varphi_{i-j+1}\cos(\varphi_{i-j+1}-\varphi_1)
\nonumber\\  
&+&\dot l_1\sum_{i=1}^N
\sum_{j=1}^{i-1}m_il_{i-j+1}\dot\varphi_{i-j+1}
\sin(\varphi_1-\varphi_{i-j+1})\nonumber\\
&+&
\sum_{i=1}^N
\sum_{j=1}^{i-1}l_{i-j+1}^2\frac{m_i}2\dot\varphi_{i-j+1}^2
+\sum_{i=1}^N\sum_{j=1}^{i-1}\sum_{k=2}^{i-j}m_il_{i-j+1}l_k
\dot\varphi_{i-j+1} \dot\varphi_k\cos(\varphi_{i-j+1}-\varphi_k)\label{tdiscr}
\end{eqnarray}
where $M=\sum_{i=1}^Nm_i$ is the total mass of the chain \footnote{Our
result agrees with that of Ref.~\cite{tomapier}, where a similar
calculation has been recently reported. Our expression of the kinetic
energy of the chain is slightly more general, since both ends of the
chain are free to move.}.

We wish now to perform the limit in which the chain of $N-1$ segments
becomes a continuous system \footnote{An exhaustive discussion about
  the passage from discrete to continuous random chain systems may be
  found in \cite{kleinertgf}.}. To this purpose, it is convenient to
consider the indices $i,j,k,\ldots$ appearing in Eq.~\ceq{tdiscr} as
discrete variables taking values in a one-dimensional lattice with $N$
points. Quantities $f_i$ carrying the index $i$ may be interpreted as
discrete functions of $i$. Their variations $\Delta f_i$ are given by:
$\Delta 
f_i=f_{i+1}-f_i$. Clearly,  $\Delta i=1$, i. e. the
spacing between two neighboring points in the lattice is 1. In order
to proceed, we rescale the distances in the lattice in such a way that
the spacing in the new lattice will be $a$. To this purpose, we
perform the transformations $i\longrightarrow s_i$,
$f_i\longrightarrow f(s_i)$ where the new variable $s_i$ has variation
$\Delta s_i=s_{i+1}-s_i=a$. The next step is to compute the kinetic
energy of Eq.~\ceq{tdiscr} in the limit 
\beq{
N\longrightarrow\infty\qquad\qquad a\longrightarrow 0\qquad\qquad
Na=L
}{contlim}
in which the product $Na$ remains finite and is equal to
the total length of the chain $L$. Clearly,
in the limit \ceq{contlim} the right hand side of Eq.~\ceq{tdiscr}
will diverge 
unless we suppose that the masses $m_i$ and the lengths $l_i$ of the
segments are going to zero in a suitable way.
Reasonable assumptions are:
\beq{l_i\longrightarrow l(s_i)=a\sigma(s_i)\qquad\qquad
m_i\longrightarrow m(s_i)=a\rho(s_i)}{ansaone}
where $\sigma(s_i)$ and $\rho(s_i)$ are respectively the
distribution of length and of mass along the chain. 
To be consistent with our settings, the distributions $\sigma(s_i)$
and $\rho(s_i)$ must be normalized as follows:
\beq{
\sum_{i=1}^N\sigma(s_i)\Delta s_i=L\qquad\qquad
\sum_{i=1}^N\rho(s_i)\Delta s_i=M
}{nordist}

While it would be interesting to study chains in which the segments
are allowed to have different lengths and the points have different masses,
for simplicity we will suppose from now on
that the length and mass distributions in
the chains are uniform, i. e.:
\beq{
\sigma(s_i)=1
\qquad\mbox{for }i=2,\ldots,N
\qquad\mbox{and}\qquad \rho(s_i)=\frac ML
\qquad\mbox{for
  }i=1,\ldots,N
}{unifass}
In the discrete case (compare with Eq.~\ceq{ansaone}) this implies that
all segments of the chain and the masses $m_i$ are equal:
\beq{l_i=a\qquad\mbox{for }i=2,\ldots,N
\qquad\mbox{and}\qquad m_i=\frac ML a\qquad\mbox{for
  }i=1,\ldots,N}{liallequal} 
At this point we are ready to pass to the continuous limit. Functions
of discrete variables will be substituted with functions of continuous
variables, while sums will be replaced with integrals according to the
following rules:
\beq{f(s_i)\longrightarrow f(s)\qquad\qquad \sum_{i=1}^N\Delta
  s_i\longrightarrow \int_0^Lds}{disccontrul}
After a few calculations one finds:
\beq{K_{disc}(t)\longrightarrow K(t)}{kdiscik}
where
\begin{eqnarray}
K(t)&=&\frac M2(l_1^2(t)\dot\varphi_1^2(t)+\dot
l_1^2(t))\nonumber\\ 
&\!\!\!\!\!\!\!\!+&\!\!\!\!\!\!\dot\varphi_1(t)l_1(t)\frac
ML\int_0^Lds
\int_0^s
du\dot\varphi(t,s-u)\cos(\varphi(t,s-u)-\varphi_1(t))\nonumber
\\
&\!\!\!\!\!\!\!\!+&\!\!\!\!\!\!\dot
l_1(t)\frac ML\int_0^Lds\int_0^sdu\dot\varphi(t,s-u)\sin(
\varphi_1(t) -\varphi(t,s-u))\nonumber\\
&\!\!\!\!\!\!\!\!+&\!\!\!\!\!\!\!\!\!\frac
ML\int_0^Lds\int_0^sdu 
\int_0^{s-u}\!\!\!dv 
\dot\varphi(t,s-u)  
\dot \varphi(t,v) \cos(\varphi(t,s-u)-\varphi(t,v))
\label{Tcont}
\end{eqnarray}
Let us note that the right hand side of Eq.~\ceq{Tcont} contains four
terms, while the original discrete version of the kinetic energy in
Eq.~\ceq{tdiscr} contained five terms. In fact, the contributions
proportional to $\dot\varphi^2_{i-j+1}$ of Eq.~\ceq{tdiscr} disappear
in the continuous limit. 
%As we will see later, this fact simplifies
%the  equation which describes the motion of
%the chain.
%and
%\beq{
%\int_0^Lds\rho_l(s)=L\qquad\qquad\int_0^Lds\rho_m(s)=M
%}{addpos}

Eq.~\ceq{Tcont} may be simplified by performing the following change
of variables: 
\beq{
u'=s-u\qquad\qquad du'=-du
}{chavar}
Using also the formula:
\beq{
\int_0^Lds\int_0^sdu'f(u')=\int_0^Lds(L-s)f(s)
}{usefor}
which is valid for any integrable function $f(s)$,
we obtain:
\begin{eqnarray}
&K(t)&=\frac M2(l_1^2(t)\dot\varphi_1^2(t)+\dot l_1^2(t))\nonumber\\
&+&\dot\varphi_1(t)l_1(t)\frac ML\int_0^Lds(L-s)
\dot\varphi(t,s)\cos(\varphi(t,s)-\varphi_1(t))\nonumber
\\
&+&\dot
l_1(t)\frac ML\int_0^Lds(L-s)\dot\varphi(t,s)\sin(
\varphi_1(t) -\varphi(t,s))\nonumber\\
&+&\!\!\!\frac ML\int_0^Lds(L-s)\int_0^sdu
\dot\varphi(t,s)  
\dot\varphi(t,u) \cos(\varphi(t,s)-\varphi(t,u))
\label{Tcontsimp}
\end{eqnarray}
%
%
%
%Remembering the definition of the length distribution in
%Eq.~\ceq{ansaone}, it is easy to realize that the first of 
%Eqs.~\ceq{unifass} implies that all segments of the chain have the
%same length. 
As a further simplification, one could fix the point
$P_1$ in some location, so that 
\beq{
\dot l_1=\dot\varphi_1=0
}{ponefixass}
Exploiting the above assumptions in Eq.~\ceq{Tcontsimp}, we find that
the Lagrangian ${\cal L}_0(t)=K(t)$ of the ideal chain is given by:
\beq{
{\cal L}_0(t)=\frac ML\int_0^Lds
(L-s)\int_0^sdu\dot\varphi(t,s)\dot\varphi(t,u)
\cos(\varphi(t,s)-\varphi(t,u))
}{hamfrecha}

What happens if, instead of polar coordinates, we choose cartesian
coordinates in order to compute the continuous limit of the kinetic
energy~\ceq{totenechadef}? 
With the help of the prescriptions given in
Eqs.~(\ref{contlim}--\ref{disccontrul}) and related comments, it is
easy to show that the Lagrangian of the ideal
chain ${\cal L}_{0,disc}=K_{disc}$ 
becomes in cartesian coordinates:
\beq{{\cal L}_0(t)=
\frac{M}{2L} \int_0^Lds(\dot x^2(t,s)+\dot y^2(t,s))
}{hzeroreal}
Of course, the fields $x(t,s)$ and $y(t,s)$ are not independent,
because they satisfy the relationship:
\beq{(x'(t,s))^2+(y'(t,s))^2=1
}{constrxy}
where $x'=\frac{\partial x}{\partial s}$ and
$y'=\frac{\partial y}{\partial s}$.
Eq.~\ceq{constrxy} is the continuous version of the constraints
\ceq{constrdiscln}.
As we see, the form of the kinetic 
energy is much simpler than that of its analogue in polar
coordinates, but the price to be paid is the cumbersome presence of
the constraints \ceq{constrxy}.
If 
$l_1$ and $\varphi_1$ are constants, 
according to the assumption of Eq.~\ceq{ponefixass}, 
one
should also add to Eqs.~\ceq{hzeroreal} and
\ceq{constrxy} 
the boundary
conditions:
\beq{
x(t,0)=l_1\cos\varphi_1\qquad\qquad y(t,0)=l_1\sin\varphi_1
}{initcond}
In this way
one end of the
chain is fixed at the given point
$(x(t,0),y(t,0))=(l_1\cos\varphi_1,l_1\sin\varphi_1)$.
 It is possible to implement other
 boundary conditions. For instance, one could ask 
that the chain forms a closed loop:
\beq{
x(t,0)=x(t,L)\qquad\qquad y(t,0)=y(t,L)
}{cloloocond} 
%It is easy to see
%that $x(t,s)$ and $y(t,s)$
%correspond in the continuous case to the discrete coordinates
%$x_n$ and $y_n$ of the point $P_n$ given by Eq.~\ceq{carpol2}.

%Written in terms of the variables
%$x(t,s)$ and $y(t,s)$, the functional ${\cal H}_0(\varphi)$
%becomes:

%where of course the fields $x(t,s)$ and $y(t,s)$ are not independent
%because 
%they
%contain only one degree of freedom, which is provided by the field
%$\varphi(t,s)$. 
%Indeed, it is not difficult to check that $x(t,s)$ and
%$y(t,s)$ 
%satisfy the constraint:

%Clearly, the kinetic energy ${\cal H}_0(x,y)$ is the continuous
%version
%of the discrete kinetic energy given in Eqs.~\ceq{totenechadef} and
%\ceq{totenechadefdef}. Moreover, the constraint \ceq{constrxy}
%corresponds in the discrete case this constraint to
%the condition:
%\beq{
%(x_i-x_{i-1})^2+(y_i-y_{i-1})^2=a^2\qquad\qquad i=2,\ldots,N
%}{constrdisc}
%which is nothing but Eq.~\ceq{constrdiscln} specialized to the present
%situation in which all the lengths $l_n$ have been set equal to $a$.

To show that the Lagrangian in cartesian coordinates \ceq{hzeroreal} 
and the Lagrangian in polar coordinates \ceq{hamfrecha} are
equivalent, it is possible to perform in Eq.~\ceq{hzeroreal}
the field transformations:
\begin{eqnarray}
x(t,s)=\int_0^sdu\cos\varphi(t,u)+l_1\cos\varphi_1
%=\frac12(\Phi(t,s)+\bar
%\Phi(t,s))%+l_1\cos\varphi_1
\nonumber\\
y(t,s)=\int_0^sdu\sin\varphi(t,u)+l_1\sin\varphi_1
%=\frac1{2i}(\Phi(t,s)-\bar
%\Phi(t,s))%+l_1\sin\varphi_1
\label{realvar} 
\end{eqnarray}
These are analogous to the discrete changes of variables of
Eqs.~\ceq{velcom1} and \ceq{velcom2} in the case in which one end of
the chain is kept fixed.
It is easy to check that, after the substitutions
\ceq{realvar} in Eq.~\ceq{hzeroreal}, one obtains exactly
Eq.~\ceq{hamfrecha} as desired.

For future convenience we introduce also
the vector notation:
\beq{
\mathbf R(t,s)=(x(t,s),y(t,s))
}{vecnot}
In this way we get for the functional ${\cal L}_0$ and the
  constraint \ceq{constrxy}
the more compact expressions:
\beq{{\cal L}_0(t)
=\frac{M}{2L} \int_0^Lds\dot \mathbf R^2(t,s)
}{dffdf}
and
\beq{
\left(\mathbf R'(t,s)\right)^2=1
}{constbfr}
Finally, Eqs. \ceq{initcond} and \ceq{cloloocond} become respectively:
\beq{\mathbf R(t,0)=(l_1\cos\varphi_1,l_1\sin\varphi_1)}
{begbb} 
\beq{
\mathbf R(t,0)=\mathbf R(t,L)
}{cloloopcondR}

It is now easy to add the
interactions. For example, let us suppose that the segments of the
chain are immersed in an  potential $V_1(\mathbf r)$ and
that there are also internal interactions associated to a two-body
potential $V_2(\mathbf r_1,\mathbf r_2)$.
In this case, the Lagrangian ${\cal L}_0$ of  Eq.~\ceq{dffdf}
generalizes to:
\beq{
{\cal L}={\cal L}_0+{\cal L}_1+{\cal L}_2
}{sgen}
where %${\cal L}_0$ has been defined in Eq.~\ceq{actionzerodef}, while
\beq{
{\cal L}_1=-\int_0^Lds V_1(\mathbf R(t,s))
}{eneext}
and
\beq{
{\cal L}_2=-\int_0^Lds_1
\int_0^Ld s_2V_{int}(
\mathbf R(t,s_1),\mathbf R(t,s_2)
)
}{eneint}
%Alternatively, if one wishes to express ${\cal A}_{ext}$ and ${\cal
%  A}_{int}$ as functionals of 
%$\Phi(t,s)$ and $\bar \Phi(t,s)$ it is possible to exploit Eqs.~\ceq{realvar}.
\section{Dynamics of a chain immersed in a thermal bath}\label{sec:three}
In this Section it will be addressed the problem of describing the
dynamics of a random chain subjected to thermodynamic fluctuations
and immersed in an environment held at
constant temperature $T$.
%We will keep such an abstract system in
% mind throughout the 
%discussion, although in principle a concrete realization of it
% could be, with some approximation, a long flexible polymer in a
% solution. 

The strategy in order to treat this
problem is to consider the discrete chain
 as a set of $N$ particles of mass $m$ performing a random walk while
subjected to the discrete constraints \ceq{constrdiscln}, which we
rewrite here 
for convenience as follows:
\beq{
\frac{\left|
\mathbf R_n(t)-\mathbf R_{n-1}(t)
\right|^2}{a^2}=1\qquad\qquad n=2,\ldots,N
}{constaconst}
%where $\mathbf R_1(t)=(l_1\cos\varphi_1,l_1\sin\varphi_1)$.
%Accordingly, we begin with a system of $N$ free
%particles of mass $m$ performing a random walk  at
%temperature $T$. 
It is additionally required that at the instants $t=t_i,t_f$ the
$n-$th particle is located respectively at the initial point $\mathbf
R_{i,n}$ and at the final point $\mathbf R_{f,n}$ for $n=1,\ldots,N$.
For simplicity, the interactions among the particles are switched off
including hydrodynamic forces \footnote{Here and in the following the
  interactions among particles of mass $m$ mediated by the motion of
  the fluid are called hydrodynamic interactions, according to the
  definition of hydrodynamic interactions given in \cite{doiedwards},
  Section
 3.6, page 66.}.
%The contribution of the
%hydrodynamic forces will be neglected, so that the motion of each
%particle will be independent of the motion of the other particles.

If one could ignore the constraints,
the probability distribution  $\Psi_N$ of the system of $N$ particles
would be: 
\beq{
\Psi_{N}=\prod_{n=1}^N\psi_n(t_f-t_i, \mathbf R_{f,n},\mathbf
R_{i,n})
}{fullnpart}
where $\psi_n$ is the probability distribution 
describing the free random walk
of the $n-$th particle. As it
is well known, $\psi_n$ satisfies the Fokker-Planck-Smoluchowski
equation 
\beq{
\frac{\partial\psi_n}{\partial(t_f-t_i)}=D\frac{\partial^2\psi_n}{\partial
\mathbf R_n^2
}
}{diffusioneq}
$D$ being the diffusion constant. Eq.~\ceq{diffusioneq} is completed
by the boundary condition:
\beq{
\psi_n(0,\mathbf R_{f,n},\mathbf
R_{i,n} )=\delta(\mathbf R_{f,n}-\mathbf
R_{i,n})
}{bdcond}
The solution of Eq.~\ceq{diffusioneq} can be written up to an irrelevant
normalization factor $A$ in the form of a path
integral:
\beq{
\psi_n=A\int_{\mathbf R_n(t_f)=\mathbf R_{f,n}\atop
\mathbf R_n(t_i)=\mathbf R_{i,n}
}
d\mathbf R_n(t)
e^{-\int_{t_i}^{t_f} \frac{\dot\mathbf R^2_n}{4D}dt}
}{psinpi} 
Substituting Eq.~\ceq{psinpi} in Eq.~\ceq{fullnpart}, the
probability distribution $\Psi_N$  becomes: 
\beq{
\Psi_N=A^N\int_{\mathbf R_1(t_f)=\mathbf R_{f,1}\atop
\mathbf R_1(t_i)=\mathbf R_{i,1}} d\mathbf R_1(t)\cdots
\int_{\mathbf R_N(t_f)=\mathbf R_{f,N}\atop
\mathbf R_N(t_i)=\mathbf R_{i,N}
}d\mathbf R_n(t)
\exp \left\{-
\sum_{n=1}^N\int_{t_i}^{t_f}\frac{\dot\mathbf R_n^2(t)}{4D}dt
\right\}
}
{postnpi}

Now we have to add to the above free random walks the
constraints \ceq{constaconst}. This will be done by inserting in the
probability 
distribution
of Eq.~\ceq{postnpi} a product of
%to derive the probability distribution
%$\Psi_{disc}$ of the two dimensional chain we start from that
%of $N$ free particles $\Psi_N$ given in
%Eq.~\ceq{postnpi} and add to it the constraints \ceq{constaconst} with
%the help of
 Dirac $\delta-$functions which enforce exactly these constraints:
\beq{
\Psi_{disc}=C
\int_{\mathbf R_1(t_f)=\mathbf R_{f,1}\atop
\mathbf R_1(t_i)=\mathbf R_{i,1}}
d\mathbf R_1(t)
\cdots
\int_{\mathbf R_N(t_f)=\mathbf R_{f,N}\atop
\mathbf R_N(t_i)=\mathbf R_{i,N}
}
d\mathbf R_n(t)
e^{-{\cal A}_{0,disc}}\prod_{n=2}^N\delta
\left(
\frac{\left|
\mathbf R_n(t)-\mathbf R_{n-1}(t)
\right|^2}{a^2}-1
\right)
}{partfunpp}
In the above equation $C$ denotes a normalization constant. Moreover,
we have introduced the functional ${\cal A}_{0,disc}$ defined as follows:
\beq{
{\cal A}_{0,disc}=\sum_{n=1}^N\int_{t_i}^{t_f}\frac{\dot \mathbf
  R_n^2(t)}{4D} dt
}{actiondisc}
%At this point we are ready to perform the limit
%\ceq{contlim} in which the chain
%becomes a continuous system. First of all, we recall that
The distribution $\Psi_{disc}$ measures the probability that a chain
starting from the initial configuration $\{\mathbf
R_{i,1},\ldots,\mathbf R_{i,N}\}$ of its segments ends up after a time
$t_f-t_i$ in the configuration  $\{\mathbf
R_{f,1},\ldots,\mathbf R_{f,N}\}$.
We note that the diffusion constant $D$ appearing in 
 \ceq{actiondisc} satisfies the relation:
\beq{
D=\mu kT
}{difcon}
where $\mu$ is the  mobility of the particle, $k$ is the Boltzmann
constant and $T$ is the temperature. 
This fact allows to rewrite the quantity
${\cal
  A}_{0,disc}$  in a form which reflects the analogy of the present
problem with a quantum mechanical problem:
\beq{
{\cal A}_{0,disc}=\frac 1{2kT\tau}\sum_{n=1}^N
\int_{t_i}^{t_f}\frac m2\dot\mathbf R_n^2(t)dt
}{azdiscresc}
In the above equation we have put
\beq{
\tau=\mu m
}{reltimedef}
The parameter $\tau$ has the dimension of a time. Indeed, $\tau$ is
the relaxation time that characterizes the rate of the decay of the
drift velocity of the particles composing the chain.
The quantity ${\cal A}_{0,disc}$ looks now like a real action of a set
 of $N$ quantum particles of mass $m$ in complex time, with
 the constant factor
 \beq{
\kappa=2kT\tau
}{hpplanck}
replacing the Planck constant $\hbar$.
This is not a surprise, because the connections between quantum
mechanics and Brownian motion are well known. Indeed, one may show
that the uncertainties in the position and momentum of a Brownian
particle are related to the constant $2mD=\kappa$ \cite{rice}.

At this point we are ready to take in the probability distribution
\ceq{partfunpp} the continuous limit \ceq{contlim}.
By introducing the rescaled variables $s_n$ as we did in
Section~\ref{sec2}, the probability distribution \ceq{partfunpp} becomes:
\begin{eqnarray}
\Psi_{disc}&=&C\prod_{n=1}^N
\int_{\mathbf R(t_f,s_n)=\mathbf R_{f}(s_n)\atop
\mathbf R(t_i,s_n)=\mathbf R_{i}(s_n)}d\mathbf R(t,s_n)
e^{
-\frac 1{2kT\tau}
\int_{t_i}^{t_f}\frac M{2L}\sum_{n=1}^N
\dot\mathbf R^2(t,s_n)\Delta s_n
}\nonumber\\
&\times&\prod_{n=2}^N\delta
\left(
\frac{\left|
\mathbf R(t,s_n)-\mathbf R(t,s_{n-1})
\right|^2}{a^2}-1
\right)
\label{partfunppint}
\end{eqnarray}
In the limit $N\longrightarrow \infty$, $a\longrightarrow 0$, $Na=L$
we obtain from $\Psi_{disc}$ the probability distribution $\Psi$
of the continuous chain:
\beq{
\Psi%[\mathbf R_f(s),\mathbf R_i(s)]
=C\int_{\mathbf R(t_f,s)=
\mathbf R_f(s)\atop
\mathbf R(t_i,s)=
\mathbf R_i(s)
}
{\cal D}\mathbf R(t,s)e^{-{\cal A}_0}\delta\left(
\left|
\frac{\partial \mathbf R(t,s)}{\partial s}
\right|^2-1
\right)
}{contparttd}
where
\beq{
{\cal A}_0=\frac
1{2kT\tau}\int_{t_i}^{t_f}dt\int_0^Lds\frac{M}{2L}\dot\mathbf R^2(t,s)
}{contacttd}
This is the desired result. 
Formally, the  normalization constant $C$ may be written as a path
integral over the initial and final configurations:
\beq{
C^{-1}=
\int{\cal D}\mathbf R_i(s){\cal D}\mathbf R_f(s)
\int_{\mathbf R(t_f,s)=
\mathbf R_f(s)\atop
\mathbf R(t_i,s)=
\mathbf R_i(s)
}
{\cal D}\mathbf R(t,s)e^{-{\cal A}_0}\delta\left(
\left|
\frac{\partial \mathbf R(t,s)}{\partial s}
\right|^2-1
\right)
}{normet}
The model described by
Eqs.~(\ref{contparttd}--\ref{contacttd}) will be called here the
generalized nonlinear sigma model due to its close resemblance to a
two-dimensional nonlinear sigma model. The most striking difference is
that the constraint $\mathbf R^2=1$ of the nonlinear sigma model has
been replaced by the condition \ceq{constbfr}, which 
contains the derivatives of the bond vectors $\mathbf R$ and it is thus
nonholonomic.

Before concluding this Section, we would like to complete the
derivation of the
 probability distribution of
Eqs.~(\ref{contparttd}--\ref{contacttd}) by discussing
 the continuous limit of the relaxation time $\tau$.
This parameter has been defined in Eq.~\ceq{reltimedef} as the product
of the mobility $\mu$ with the mass $m$ of the beads.
When the distance $a$ between the beads goes to zero, $m$ goes to zero
as well according to Eq.~\ceq{liallequal}. On the other side, 
with
decreasing values of $a$, two contiguous beads will become closer and
closer until they  will eventually merge one into
another.
To avoid this unphysical situation, one should add to the continuous
limit \ceq{contlim} the requirement that the dimensions of the
beads  vanish together with $a$. Supposing for instance  that the 
beads are circles of radius $\rho$, for dimensional
reasons one is lead to put $\rho=c a$, where $c$ is a dimensionless
proportionality factor. As it is intuitive, 
when the size of a bead decreases, its
mobility 
$\mu$ increases.
The increasing of $\mu$ compensates the vanishing of $m$,
so that the product $\tau = \mu m$
remains finite.
 This fact can be verified rigorously in three
dimensions using the well known Stokes formula of the mobility.

\section{The classical solutions of the generalized nonlinear sigma
  model}

%In this Section we consider the model of the fluctuations of a
%random chain given in Eqs.~\ceq{contparttd} and \ceq{contacttd}. This
%model closely resembles a two dimensional nonlinear sigma model, apart
%from the fact that in the nonlinear sigma model
%the constraint
%imposed with the help of a Dirac delta function
% is  $\mathbf R^2=1$.
%In the present case
%the constraint imposed in Eq.~\ceq{contparttd} is that of 
%Eq.~\ceq{constbfr} and it is non-holonomic.

To get rid of the delta function appearing in the formulation of the
generalized sigma model of
Eqs.~\ceq{contparttd} and \ceq{contacttd}, it will be convenient to
  introduce a Lagrange multiplier $\lambda=\lambda(t,s)$ and
 to use the Fourier representation of the Dirac delta function.
Moreover,
we add a coupling of the bond vectors $\mathbf R(t,s)$ 
with
an external source $\mathbf J(t,s)$.
In this way the generating functional $\Psi[J]$ of the
correlation functions of the bond vectors may be written in the form:
\beq{
\Psi[J]=\int%_{\mathbf R(t_f,s)=\mathbf R_f(s)\atop
%\mathbf R(t_i,s)=\mathbf R_i(s)
%}
{\cal D}\mathbf R{\cal D}\mathbf \lambda \exp\left\{-
\int_{t_i}^{t_f}dt\int_0^Lds\left[
\frac{M}{2L\kappa}\dot\mathbf R^2+i\lambda\left(
{\mathbf R'}^2-1
\right)
+\frac 1\kappa\mathbf J\cdot \mathbf R\right]
\right\}
}{pathlagmult}
%where we have put:
%\beq{
%\int d^2\xi=\int_{t_i}^{t_f}dt\int_0^Lds\qquad\mbox{and}\qquad
%\xi_1=t,\xi_2=s 
%}{posione}
% In the above equation we have added an external source $\mathbf J$
% for the field $\mathbf R$. 
 Let us note that, 
for the sake of generality,
in Eq.~\ceq{pathlagmult}
no boundary conditions
 for the relevant fields have been specified.
 As it stands, Eq.~\ceq{pathlagmult} 
 could be applied both to open
 or closed chains. Moreover, in the case of
 open chains both possibilities of free or fixed
end points are allowed.
It turns out that
the degree of complexity of  the computation of
 $\Psi[J]$ strongly depends on
 the choice of boundary
 conditions.

Let's now derive  the solutions $\mathbf R_{cl}$ and
 $\lambda_{cl}$ of the classical equations of motion associated with
 the generating functional $\Psi[J]$ of Eq.~\ceq{pathlagmult}:
\begin{eqnarray}
\frac{M}{L}\frac{\partial^2\mathbf R}{\partial
  t^2}&=&\mathbf J\label{cleqone}\\ 
\frac{\partial \lambda}{\partial s}\frac{\partial
  \mathbf R}{\partial s}+\lambda\frac{\partial^2\mathbf R}{\partial s^2}&=&0
  \label{cleqtwo} \\
\left|\frac{\partial \mathbf R}{\partial
    s}\right|^2&=&1\label{cleqthree}  
\end{eqnarray}
It is easy to see that, due to the constraint \ceq{cleqthree}, the current
$\mathbf J$
must depend only on the variable $t$,
i.~e. $\mathbf J(t,s)=\mathbf J(t)$. Always for the same reason, it is
possible 
to check that Eq.~\ceq{cleqtwo} is inconsistent unless
$\lambda_{cl}=const\ne 0$ or $\lambda_{cl}=0$. 
Depending on the fact that $\lambda_{cl}$ is zero or not,
 one finds that the
relevant solutions of
Eqs.~(\ref{cleqone}--\ref{cleqthree}) may be divided into two groups,
which we call here solutions of type A and of type B. The solutions of
type A are characterized by the condition
$\lambda_{cl}=0$.
They are displayed in Table~\ref{tableone}.
\begin{table}{\bf Type A solutions}
\begin{flushleft}
\begin{eqnarray}
\lambda_{cl}&=&0\label{alambda}\\
\mathbf R_{cl}(t,s)&=&\mathbf R_{0,A}+\mathbf Vt+\mathbf
R_{1,A}(t)+\mathbf 
R_{2,A}(s)\label{aerre} 
\end{eqnarray}
where $\mathbf R_{0,A}$ and $\mathbf V$ are constant vectors,
\beq{
\mathbf R_{1,A}(t)=\int_{t_i}^{t_f}dt'G(t,t')\mathbf J(t')
}{erreone}
and
\beq{
\mathbf R_{2,A}(s)=\int_0^sdu(\cos\varphi(u),\sin\varphi(u))
}{erretwo}
Here $\varphi(u)$ is an arbitrary function of $u$, while in
Eq.~\ceq{erreone} $G(t,t')$ denotes the Green function which solves
the differential equation:
\beq{
\frac M{L}\frac{\partial^2G(t,t')}{\partial t^2}=-\delta(t-t')
}{gttpdef}
Type A solutions admit closed chain configurations. In that
case, the functions $\varphi(s)$ must satisfy the additional
periodicity condition:
\beq{\varphi(s+L)=\varphi(s)}{pppre}

\end{flushleft}
\caption{Solutions of type A of the classical equations of motion
  (\ref{cleqone}--\ref{cleqthree}).} 
\label{tableone}
\end{table}
Besides the classical solutions of type A, there are also the solutions of
type B listed in Table~\ref{tabletwo}.
\begin{table}
{\bf Type B solutions}
\begin{flushleft}
\begin{eqnarray}
\lambda_{cl}&=&\mbox{const}\ne0\label{lambdab}\\
\mathbf R_{cl}(t,s)&=&\mathbf R_{0,B}+\mathbf Vt+\mathbf R_{1,B}(t)+\mathbf
R_{2,B}(s)\label{erreb} 
\end{eqnarray}
where $\mathbf R_{0,B}$ and $\mathbf V$ are  constants vectors,
\beq{
\mathbf R_{1,B}(t)=\int_{t_i}^{t_f}dt'G(t,t')\mathbf J(t')
}{errebone}
and
\beq{
\mathbf R_{2,B}(s)=(0,s)
}{errebtwo}
In the absence of the external source $\mathbf J(t)$, this kind of
solutions corresponds 
to a configuration in which the chain is stretched along the
$y-$axis with one end in the point $\mathbf R_{0,B}$
and the other end in the point $\mathbf R_{0,B}+(0,L)$. No closed loop
configuration is allowed.
Of course, the stretched chain may be oriented in a different way by
means of a rotation.\\[1ex]
\end{flushleft}
\caption{Solutions of type B of the classical equations of motion
  (\ref{cleqone}--\ref{cleqthree}).}
\label{tabletwo}
\end{table}

We would like to stress the fact that, if there are no external
currents, both type A and B of classical solutions are static,
i.~e. they do not depend on time apart from the rigid translations of
the whole chain with constant velocity $\mathbf V$. This result
is confirmed if one studies the equations of motion corresponding to
the Lagrangian \ceq{hamfrecha} obtained using polar coordinates.
The only allowed classical solutions for the field
$\varphi(t,s)$ are in fact time independent.

\section{Computation of the generating functional $\Psi[J]$ in the
  semi-classical approximation}
The exact computation of
 $\Psi[J]$ is a formidable problem despite the simplicity of the
action of the generalized nonlinear sigma model. 
One of the main difficulty is the presence of the nonholonomic
constraint $|\mathbf R'|^2=1$ in the path integral \ceq{contparttd}.
In principle, this cumbersome condition may be easily eliminated
by introducing a scalar field 
$\mathbf \varphi(t,s)$ and performing the formal substitutions
of Eq.~\ceq{realvar}
\beq{
\mathbf R(t,s)=\int_0^sdu(\cos\varphi(t,u),\sin\varphi(t,u))
}{formsubst}
Here we have assumed for simplicity that
the chain
has one fixed end in the origin of the
coordinates, so that
$\mathbf R(t,0)=(0,0)$ in agreement with Eq.~\ceq{formsubst}.
To show that after the field transformation \ceq{formsubst}
the constraint disappears from the path integral \ceq{contparttd}, we use
the following relation which will be proved in Appendix~A for a
generic functional $f(\mathbf R(t,s))$:
\beq{
\int{\cal D}\mathbf R(t,s)f(\mathbf R(t,s))\delta(|\mathbf R'|^2-1)=
{\cal N}\int{\cal D}\varphi(t,s)f(\int_0^sdu(\cos\varphi(t,u),
\sin\varphi (t,u)))
}{funcid}
where ${\cal N}$ is an irrelevant constant. In our particular case, in
which:
\beq{
f(\mathbf R(t,s))=\exp\left[-\int_{t_i}^{t_f}dt\int_0^Lds\frac{M}{2\kappa
  L}\dot\mathbf R^2\right]
}{arbfundef}
one obtains from Eq.~\ceq{funcid}:
\beq{\Psi={\cal N}\int{\cal
   D }\varphi(t,s)e^{-\frac 1\kappa\int_{t_i}^{t_f}dt{\cal L}_0}} {dfhjsfd}
with ${\cal L}_0$ being the Lagrangian defined in Eq.~\ceq{hamfrecha}.
As we see, the  Dirac delta function
 with the constraint is no longer present in
 the path integral \ceq{dfhjsfd},
% but 
%the nonlinear Lagrangian ${\cal L}_0$ leads to a theory which
%is too difficult to be treated analytically.
but the Lagrangian  of Eq.~\ceq{hamfrecha}
is both nonlocal and nonlinear.

In the following, we will  stick to cartesian coordinates 
 limiting ourselves  to study
 small gaussian fluctuations of the field $\mathbf R(t,s)$
around the classical solutions
derived in the previous Section.
To this purpose,  in
Eq.~\ceq{pathlagmult} 
we split both fields $\mathbf R(t,s)$ and $\lambda(t,s)$ into
classical contributions $\mathbf R_{cl}, \lambda_{cl}$ and 
statistical corrections $\delta\mathbf R,\delta \lambda$:
\begin{eqnarray}
\mathbf R(t,s)&=&\mathbf R_{cl}(t,s)+\kappa^{\frac 12}\delta \mathbf
R(t,s)\label{splitR} \\
\lambda(t,s)&=&\lambda_{cl}+\kappa^{\frac 12}\delta\lambda(t,s)\label{splitl}
\end{eqnarray}
Moreover, it will also be convenient to split the external source
$\mathbf J$ in an analogous way:
\beq{
\mathbf J(t,s)=\mathbf J_{cl}(t)+\kappa^{\frac 12}\delta\mathbf
J(t,s)
}{splits}
where $\mathbf J_{cl}(t)$ denotes the current depending only on the time
$t$ appearing in Eqs.~\ceq{erreone}
and \ceq{errebone}.
Due to the fact that the Lagrange multiplier $\lambda$ is just an
auxiliary field, it is possible to choose for its variation
$\delta\lambda$ trivial boundary conditions at the initial and final
instants:
\beq{
\delta\lambda(t_i,s)=\delta\lambda(t_f,s)=0
}{tribouconlam}
The boundary conditions of $\delta\mathbf R(t,s)$ will be fixed later.

At this point, we expand the action
\beq{{
\cal A
}
=
\int_{t_i}^{t_f}dt\int_0^Lds\left[
\frac{M}{2L\kappa}\dot\mathbf R^2+i\lambda\left(
{\mathbf R'}^2-1
\right)
+\frac 1\kappa\mathbf J\cdot \mathbf R\right]
}{actionbeq}
appearing in the path integral \ceq{pathlagmult} with
respect to the 
quantities $\delta \mathbf R,\delta\lambda$ and $\delta\mathbf
J$. Since the latter are supposed to be  small corrections of the
dominating classical 
solutions, we stop the expansion at the second order:
\beq{
{\cal A}={\cal A}_{cl}+\delta{\cal A}^{(1)}+\delta{\cal A}^{(2)}
}{expafor}
At the zeroth order we have:
\beq{
{\cal A}_{cl}=\frac 1\kappa\int_{t_i}^{t_f}dt\int_0^Lds\left[
\frac{M}{2L}\dot\mathbf R^2_{cl}+\mathbf J_{cl}\cdot\mathbf R_{cl}
\right]
}{acldef}
This is just the action ${\cal A}$  in which the fields have been
replaced by their classical configurations, which may be either of
type A or of type 
B. In both cases,
the term $\lambda_{cl}(\mathbf R^{\prime\, 2}-1)$, which in
principle should be present in Eq.~\ceq{acldef}, has been
omitted because   Eq.~\ceq{cleqthree} forces it to vanish identically.
Let's now compute the first order contribution $\delta{\cal
  A}^{(1)}$. 
Usually,  first order contributions vanish
after exploiting the classical equations of
motion. In our case this is in general not true. The reason
is that,
due to the nontrivial boundary conditions satisfied by $\mathbf
R(t,s)$, nonzero boundary terms may still appear in $\delta {\cal A}^{(1)}$.
Despite this fact, it is possible to show that $\delta{\cal A}^{(1)}$
vanishes at least in the following two situations:
\begin{enumerate}
\item Closed chains satisfying the boundary conditions
  \ceq{cloloopcondR}.
\item Open chains in which both ends are fixed, so that besides the
  condition \ceq{begbb} also the following one is valid:
$\mathbf R(t,L)=(x_{0,L},y_{0,L})$, where $x_{0,L}$ and $y_{0,L}$ are
  constants. 
\end{enumerate}
Assuming that one
of the above two conditions is verified, it is possible to put:
\beq{
\delta {\cal A}^{(1)}=0
}{vandeaone}
Thus, we are left only with the computation of the
second order corrections $\delta{\cal
  A}^{(2)}$. After simple 
calculations, one finds:
\beq{
\delta{\cal A}^{(2)}=\int_{t_i}^{t_f}dt\int_0^Lds\left[
\frac M{2L}\delta\dot\mathbf R\cdot\delta\dot\mathbf
R+i\kappa\lambda_{cl}(\delta\mathbf R') ^2+2i\kappa(\mathbf
R_{cl}'\cdot\delta\mathbf R')\delta\lambda+\delta \mathbf
J\cdot\delta\mathbf R
\right]
}{dstwodef}
Putting together all the above results, 
it is possible to conclude that,
within the present gaussian approximation, the expression
of $\Psi[J]$ reduces to:
\beq{
\Psi[J]=e^{{\cal A}_{cl}}Z[J]
}{probfungauss}
where
\beq{
Z[J]=\int{\cal D\delta \mathbf R}{\cal D}\delta\lambda e^{-\delta {\cal
    A}^{(2)}} 
}{partfungauss}
and ${\cal A}_{cl}$ and $\delta {\cal
    A}^{(2)}$ are respectively given in Eqs.~\ceq{acldef} and
\ceq{dstwodef}.

From this point on we will consider only chain configurations which
 are closed, so that both classical
 configurations $\mathbf R_{cl}$ and their statistical corrections
 $\delta\mathbf R$ must
satisfy  the boundary conditions in $s$ of
 Eq.~\ceq{cloloopcondR}.
In the Lagrange multiplier sector, besides the trivial boundary
 conditions in time of Eq.~\ceq{tribouconlam}, we require also the
 following ones with respect to the variable~$s$:
\beq{
\delta\lambda(t,0)=\delta\lambda(t,L)
}{lambdatriv}
The case of closed chains is particularly
 interesting because under this 
 assumption
the  path integral appearing in the right hand side of
 Eq.~\ceq{partfungauss}
 may be rewritten in such a
 way that it closely resembles the generating functional of a field theory
 of a one-dimensional system 
at finite
 temperature. In this field theory the coordinate $s$  plays the role
 of time
 while the real time $t$ becomes
  the spatial 
 coordinate of the one-dimensional space. 
We  still need to specify the boundary
conditions with respect to the time for the fields $\delta \mathbf R$. 
We fix them in
such a way that the gaussian path integration over these fields in the
generating functional of Eq.~\ceq{partfungauss} will be as simple as
possible. To this purpose, reasonable choices are the following:
\begin{description}
\item[\rm Dirichlet--Dirichlet boundary conditions]
\beq{
\delta\mathbf R(t_i,s)=0\qquad\qquad\delta\mathbf R(t_f,s)=0
}{dirdir}
\item[\rm Dirichlet--Neumann boundary conditions]
\beq{
\delta\mathbf R(t_i,s)=0\qquad\qquad\left.
\frac{\partial\mathbf R(t,s)}{\partial
  t}\right|_{t=t_f} =0
}{dirneu}
\item[\rm Neumann--Dirichlet boundary conditions]
\beq{
\left.\frac{\partial\delta\mathbf R(t,s)}{\partial
  t}\right|_{t=t_i} =0\qquad\qquad\delta\mathbf R(t_f,s)=0
}{neudir}
\item [\rm Neumann--Neumann boundary conditions]
\beq{
\left.\frac{\partial\delta\mathbf R(t,s)}{\partial
  t}\right|_{t=t_i} =0\qquad\qquad
\left.\frac{\partial\delta\mathbf R(t,s)}{\partial
  t}\right|_{t=t_f} =0
}{neuneu}
\end{description}
At this point we are ready to perform the Gaussian integrations over
the fields $\delta\mathbf R$ in the generating functional $Z[J]$ given in
Eq.~\ceq{partfungauss}. 
After some  integrations by parts, which do not
produce boundary terms thanks to the 
boundary conditions \ceq{cloloopcondR} and \ceq{lambdatriv}, one finds:
\beq{
Z[J]=C_1\int{\cal D}\delta\lambda e^{-S(\delta\lambda)}
}{zc1dl}
with
\begin{eqnarray}
S(\delta\lambda)&=&\frac 1{2}\int_0^L ds \int_{t_i}^{t_f}dt
dt'G(t,t') \left[
-4\kappa^2\delta\lambda(t,s)\frac{\partial
  \mathbf R_{cl}(s)}{\partial s}\cdot\frac{\partial^2}{\partial s^2}\left(
\delta\lambda(t',s)\frac{\partial\mathbf R_{cl}(s)}{\partial s}
\right)-
\right.\nonumber\\
&-&2i\kappa\delta\lambda(t,s)\frac{\partial\mathbf R_{cl}(s)}{\partial
  s}\cdot\frac{\partial \mathbf \delta \mathbf J(t',s)}{\partial s}
-2i\kappa\delta\lambda(t',s)\frac{\partial\mathbf R_{cl}(s)}
{\partial s}\cdot\frac{\partial
  \delta \mathbf J(t,s)}{\partial s}+\nonumber\\
&+&
\left.\phantom{\frac{\partial^a_c(\tau)}{\partial\tau}(\tau)}
\!\!\!\!\!\!\!\!\!\!\!\!\!\!\!\!\!\!\!\!\!\!\!\!
\delta\mathbf J(t,s)\cdot \delta \mathbf J(t',  
s)\right] \label{actslam}
\end{eqnarray}
and
\beq{
C_1=\int{\cal D}\delta\mathbf R e^{-\frac{M}{2L}\int_0^L
  d s\int_{t_i}^{t_f} 
dt\left(\frac{\partial\delta\mathbf R}{\partial t}\right)^2%\cdot
%\frac{\partial\mathbf R}{\partial t}
}
}{conedef}
The symbol $G(t,t')$ denotes the propagator
\ceq{gttpdef} computed taking into account one of the boundary
conditions defined in Eqs.~(\ref{dirdir}--\ref{neuneu}) \cite{gfl}. We
have thus 
four possibilities:
\begin{description}
\item[\rm Dirichlet--Dirichlet boundary conditions]
\beq{
G(t,t')=-\frac LM \theta(t'-t) (t-t_i)
\frac{(t'-t_f)}{t_f-t_i}
-\frac LM \theta(t-t')
(t'-t_i)\frac{t-t_f}{t_f-t_i}
}{propdirdir}
\item[\rm Dirichlet--Neumann boundary conditions]
\beq{
G(t,t')=-\frac LM (t_i-t')\theta(t-t')-\frac LM(t_i-t)\theta(t'-t)
}{propdirneu}
\item[\rm Neumann--Dirichlet boundary conditions]
\beq{
G(t,t')=\frac LM(t_f-t')\theta(t'-t)+\frac LM(t_f-t)\theta(t-t')
}{propneudir}
\item[\rm Neumann--Neumann boundary conditions]
\beq{
G(t,t')=\frac LM
\left[
t\theta(t'-t)+t'\theta(t-t')
\right]
+
\frac LM
\frac{
\left[(t-t_f)^2+(t'-t_f)^2
\right]}{2(t_f-t_i)}
}{propneuneu}
\end{description}
In Eqs.~(\ref{propdirdir}--\ref{propneuneu}) 
the function
$\theta(t)$ is the usual $\theta-$function of
Heaviside. Let us note that the function $G(t,t')$ of
Eq.~\ceq{propneuneu} is actually a pseudo Green function, which
satisfies the equation:
\beq{
\frac ML\frac{\partial^2G(t,t')}{\partial t^2}=-\delta(t-t')+\frac 1{t_f-t_i}
}{pseudogreen}
instead of Eq.~\ceq{gttpdef}.
This is due to the fact that, in the case of Neumann--Neumann boundary
conditions, one should project out the  constant solution of the
eigenvalue equation associated to the operator
$\frac ML\frac{\partial^2}{\partial t^2}$.

We remark that in the action \ceq{actslam} the classical
conformations appear only in the derivatives
$\frac{\partial\mathbf R_{cl}}{\partial s}$, which
coincide with the derivatives
of the vectors $\mathbf R_{2,A}(s)$ defined in
Eq.~\ceq{erretwo}. In components:
\beq{\frac{\partial\mathbf R_{cl}}{\partial s}
=(\cos\varphi(s),\sin\varphi(s))
}{univec}
From Eq.~\ceq{univec} it is clear
  that the vector 
 $\frac{\partial\mathbf R_{cl}}{\partial s}$ has the meaning of
the unitary vector
which is tangent to the classical trajectories. % $\psi^a_c(x,\tau)$.
       It is  therefore convenient to decompose all vectors appearing in the
 action 
$S(\delta\lambda)$ in components which are normal or tangent to
$\mathbf R_{cl}(s)$.
After some algebra, one obtains in this way an
expression of $Z[J]$ in which
%\beq{
%Z[j]=\int{\cal D}\Lambda e^{-S(\Lambda)}
%}{znew}
%where 
 now $S(\delta\lambda)$ takes the simplified form:
\begin{eqnarray}
S(\delta\lambda)&=&\frac 12\int_0^L ds\int_{t_i}^{t_f}dtdt'
G(t,t')\left[
-4\kappa^2\delta\lambda(t,s)\left(
\frac{\partial ^2}{\partial s^2}-(\varphi'(s))^2
\right)\delta\lambda(t',s)\right.+\nonumber\\
&+&\left.2i\kappa\delta\lambda(t,s)\delta
J_T(t',s)+2i\kappa\delta\lambda(t',s)
\delta J_T(t,s)  
+ \delta\mathbf
 J(t,s)\cdot
\delta\mathbf  J(t',s)\phantom{\frac{\partial ^2}{\partial\tau^2}}
\!\!\!\!\!\!\!\!\!\!
\right]\label{simpactsdl}
\end{eqnarray}
Here we have introduced the {\it tangential} component $\delta J_T(t,s)$
of the current $\mathbf \delta \mathbf J(t,s)$:
\beq{
\delta J_T(t,s)=\partial_s\mathbf R_{cl}(s)\cdot
\partial_s \mathbf \delta\mathbf J(t,s)
}{tancur}
At this point we are ready to eliminate the auxiliary field $\delta\lambda$
from the functional of Eq.~\ceq{zc1dl}, where now the action $S(\delta
\lambda)$
is defined in Eq.~\ceq{simpactsdl}.
To this purpose, one needs to perform a  gaussian integration,
which produces the result:
\beq{Z[J]=C_1C_2
e^{
-\frac 12\int_0^L ds ds'\int_{t_i}^{t_f}dtdt'G(t,t')\left[
K(s,s')\delta J_T(t,s)\delta J_T(t',s')+\frac 1 L\delta\mathbf J(t,s)
\cdot\mathbf J(t',s)
\right]
}
}{zfin}

In the above equation $K(s,s')$ denotes the Green function which
satisfies the 
relation:
\beq{
\left[
\frac{\partial^2}{\partial s^2}-(\varphi'(s))^2
\right]K(s,s')=-\delta(s,s')
}{gretwoeqdef}
while
\beq{
C_2=\int{\cal D}\delta\lambda e^{\int_0^L ds
\int_{t_i}^{t_f}dtdt'G(t,t')\left[
2\kappa^2\delta\lambda(t,s)\left(
\frac{\partial^2}{\partial s^2}-(\varphi'(s))^2
\right)\delta \lambda(t',s)
\right]
}
}{ctwodef}
Putting all together we
obtain the expression of the generating 
functional $\Psi[J]$ in its final form:
\begin{eqnarray}
\Psi[J]&=&e^{{\cal A}_{cl}}C_1C_2
\exp\left[-\frac 1L
\int_0^L ds\int_{t_i}^{t_f}dtdt'
G(t,t')\delta \mathbf
J(t,s)\cdot\delta\mathbf J(t',s)
\right]
\nonumber\\
&\times&\exp\left[
-\frac 12\int_0^L dsds'\int_{t_i}^{t_f}dtdt'G(t,t')
K(s,s')\delta J_T(t',s'))\delta J_T(t,s)
\right]\label{finforgenfun}
\end{eqnarray}
The right hand side of Eq.~\ceq{finforgenfun} displays the
 asymmetry in the propagation of transverse and longitudinal modes.

The differential equation satisfied by the Green function
$K(s,s')$ of Eq.~\ceq{gretwoeqdef}
cannot
be solved 
analytically for any given function $\varphi(s)$.
Here we discuss just the case in which the background classical
solution corresponds to a chain with the configuration of a circle, i.~e.:
\beq{
\mathbf R_{cl}^{circle}(s)=\frac L{2\pi}\left(
\cos\frac{2\pi s}{L},\sin\frac{2\pi s}{L}
\right)
}{Rclcirc}
The radius of the circle is $\frac
L{2\pi}$, so 
that the total length of the chain is $L$ as desired. Comparing
Eq.~\ceq{Rclcirc} with Eq.~\ceq{erretwo}, it is clear that for this
conformation $\varphi(s)=\frac {2\pi}Ls$. Substituting this expression
of $\varphi(s)$ in Eq.~\ceq{gretwoeqdef}, it turns out that the Green function
$K(s,s')$ satisfies the relation:
\beq{
\left(
\frac{\partial^2}{\partial s^2}-\frac{4\pi^2}{L^2}
\right)K(s,s')=-\delta(s-s')
}{relkcir}
The solution of the above equation corresponding to the boundary
conditions \ceq{lambdatriv} is:
\beq{
K(s,s')=
\frac{\sinh\left[
\frac{2\pi}L(L-s')
\right]\sinh\frac{2\pi}Ls}{\frac {2\pi}L\sinh 2\pi}\theta(s'-s)+
\frac{\sinh\left[
\frac{2\pi}L(L-s)
\right]\sinh\frac{2\pi}Ls'}{\frac {2\pi}L\sinh 2\pi}\theta(s-s')
}{kcircleder}

In the limit $\mathbf J(t,s)=0$, we obtain 
from the generating functional of Eq.~\ceq{finforgenfun}
the expression of the
probability distribution $\Psi$ of
Eqs.~(\ref{contparttd}--\ref{contacttd}) in the semiclassical approximation:
\beq{
\Psi=e^{{\cal A}_{cl}(\mathbf J_{cl}=0)}C_1C_2
}{nomade}
Remembering the respective definitions of the constants $C_1$ and
$C_2$ of Eqs.~\ceq{conedef} and \ceq{ctwodef}, together with the form
\ceq{acldef} of the classical action, $\Psi$ may be explicitly written
as follows:
\begin{eqnarray}
\Psi&=&\exp\left\{-\frac 1\kappa\int_{t_i}^{t_f}dt
\int_0^Lds\left[
\frac M{2L}\dot \mathbf R^2_{cl}
\right]
\right\}
\int{\cal D}\delta\mathbf R \exp\left[-{\displaystyle
\frac M{2L}\int_{t_i}^{t_f}dt
\int_0^Lds\left(\delta\dot\mathbf R\right)^2
}\right]\nonumber\\
\!\!\!\!\!\!
&\!\!\!\!\!\!
\times&\!\!\!\!\!\!\int{\cal D}\delta\lambda\exp\left\{
\int_{t_i}^{t_f}dtdt'\int_0^LdsG(t,t')\left[
2\kappa^2\delta\lambda(t,s)\left({\textstyle
\frac{\partial^2}{\partial s^2}-(\varphi'(s))^2
}\right)\delta\lambda(t',s)
\right]
\right\}\label{finprofun}
\end{eqnarray}

\section{Physical Interpretation of the obtained results}

The classical equations of motion (\ref{cleqone}--\ref{cleqthree})
admit only static solutions in which the conformation of the chain is
fixed. Only  rigid translations of the chain as a whole
with constant velocity $\mathbf V$ are allowed. Apart from these rigid
translations, the time dependence of the classical solutions $\mathbf
R_{cl}(t,s)$  in Eqs.~\ceq{aerre} and \ceq{erreb} is just an artifact
of the presence of the classical current $\mathbf J(t)\equiv\mathbf
J_{cl}(t)$ which is fictitious and may thus be set to zero without
any loss of generality. In the following discussion, it will be assumed that
both $\mathbf J_{cl}(t)$ and $\mathbf V$ are zero.
The absence of any relevant dynamics in the classical solutions is
somewhat surprising. We can only suggest that this absence could be
related to the fact that, after 
performing the continuous limit, the fourth term appearing in the left
hand side of Eq.~\ceq{tdiscr} vanishes identically. 
This term is important for the chain dynamics since it contains second time
derivatives of the angles $\varphi(t,s)$.
In some sense, the
continuous chain is simpler that its discrete counterpart, in which
this term is present.

\begin{figure}[bhpt]
\includegraphics[width=6cm]{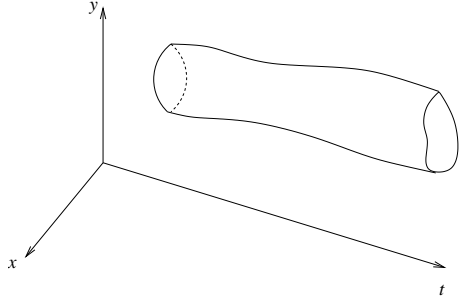}
\caption{During its motion, a closed chain spans in the $x,y,t$ a
  surface which has the topology of a cylinder.}
\label{cylclocon}
\end{figure}

Less trivial is the treatment of the fluctuations of the chain at
constant temperature $T$. 
The nonlinear sigma model given in Eq.~\ceq{pathlagmult}
or, alternatively, its formulation without the Lagrange multiplier of
Eqs.~(\ref{contparttd}--\ref{contacttd}), describe the fluctuations of
a chain of $N$ segments of constant length $a$ in the continuous limit
\ceq{contlim}.
During its motion in the time interval $[t_i, t_f]$ the chain
spans a two-dimensional surface which, in the case of a closed
conformation, has the topology of a cylinder in the $x,y,t$ space, see
Fig.~\ref{cylclocon}.
If the chain is open, instead, the topology of the cylinder should be
replaced with that of a strip. 

We recall that in deriving the GNLSM of
Eqs.~(\ref{contparttd}--\ref{contacttd}) the contribution of the
hydrodynamic interactions has been neglected. This limits the validity
of this model to the following cases:
\begin{description}
\item[\rm a)] The viscosity of the fluid is large, so that the motion
  of the particles composing the chain is slow. Thus the velocity
  field created by each particle is too weak to influence the motion
  of other particles.
\item[\rm b)] The temperature is low, so that once again the motion of
  the chain is slow.
\item[\rm c)] The conformation of the chain is relatively straight
  because there is some energy cost when the chain is being bent. To
  this purpose, however, one should introduce the stiffness at the
  joints between the segments.
\end{description}

%In this work we have computed the generating
%functional $\Psi[J]$ of the correlation functions:
%\beq{
%G_{i_1\ldots,i_M}(t_1,s_1,\ldots,t_M,s_M)=\left\langle
%R_{i_1}(t_1,s_1)\ldots R_{i_M}(t_M,s_M)
%\right\rangle
%}{genfuncorfun}
%where the indices $i_1,\ldots,i_M=1,2$ label the components of the
%vector $\mathbf R(t,s)$.

On the other side, the semiclassical approximation used in order to derive
the generating functional
 $\Psi[J]$ of Eq.~\ceq{finprofun} is
valid in the case in which the parameter $\kappa$ defined in
Eq.~\ceq{hpplanck} is small. This parameter depends essentially on the
temperature $T$ and on the relaxation time $\tau$.
Since
$\tau$ is inversely proportional to the viscosity in the limit of low
Reynolds number, 
it is reasonable to assume that the semiclassical approach can be
applied to a cold isolated chain or to an isolated chain in a very
viscous solution. These situations correspond respectively to the
points b) and a) mentioned above.

Both the generating functional $\Psi[J]$ of Eq.~\ceq{finforgenfun}
and the probability distribution of Eq.~\ceq{finprofun} have been
computed in the case of closed chains, whose conformations are
subjected to the boundary conditions (\ref{dirdir}--\ref{neuneu}).
The physical meaning of these  boundary conditions 
may be summarized as follows.
%The four possibilities stated in Eqs.~(\ref{dirdir}--\ref{neuneu})
%correspond to the following physical situations:
 \begin{description}
\item[\rm Dirichlet--Dirichlet boundary conditions:] 
In this case $\mathbf R_i(s)=\mathbf R_f(s)=\mathbf R_{cl}(s)$, where
$\mathbf R_{cl}(s)$ is a given static solution of the classical
equations of motion.
The probability
  that, starting from the conformation $\mathbf R_i(s)$, the
  fluctuating chain ends up at the time $t_f$ in the  same
  conformation, is proportional up to a normalization
  constant to the probability distribution $\Psi$ of
  Eq.~\ceq{finprofun}, in which the Green function $G(t,t')$ is that
  of Eq.~\ceq{propdirdir}.
\item[\rm Dirichlet--Neumann boundary conditions:] In this case the
  probability 
  distribution $\Psi$ 
of Eq.~\ceq{finprofun}
is proportional to the probability that a closed
  chain starting from a classical static  conformation $\mathbf R_i(s)=\mathbf
  R_{cl}(s)$ at the time $t_i$ ends up at the instant $t_f$ in an
  arbitrary conformation characterized by the fact that the
  velocities of each segment 
  composing the chain is zero. $\Psi$ must be computed choosing the
  Green function $G(t,t')$ defined in Eq.~\ceq{propdirneu}.
\item[\rm Neumann-Dirichlet:] Here the segments of the chain have zero
  velocity at the beginning, but the conformation of the chain is
  otherwise arbitrary. At the time $t_f$ the chain is found in a given
  static classical conformation, i.~e. $\mathbf R_f(s)=\mathbf R_{cl}(s)$. The
  probability for this to happen is obtained
  after substituting in  Eq.~\ceq{finprofun} the Green function
  $G(t,t')$ of 
  Eq.~\ceq{propneudir}. 
\item[\rm Neumann-Neumann:] This is the situation in which the
  conformation of the chain at the initial and final times $t_i$ and
  $t_f$ are not specified, but the velocities of all the segments
  composing the chains must be zero. The relevant Green function
  $G(t,t')$ to be inserted in the probability distribution $\Psi$ is
  that of Eq.~\ceq{propneuneu}.
\end{description}
By choosing Neumann--Dirichlet boundary conditions one may check for
instance if, starting from any static conformation, there is a particular
conformation in which it is very likely that
the chain will be found after a certain time
$t_f-t_i$.
The stability of a given conformation with respect to the
thermodynamic fluctuations which attempt to reshape the chain can be
tested by choosing Dirichlet--Dirichlet boundary conditions. 
In principle, it is also possible to study other types of boundary
conditions than those considered here, provided they do not give rise
to unwanted boundary terms in the action of the generating functional
$\Psi[J]$.

To conclude this Section, it is interesting to see how the original
constraint \ceq{constbfr} is realized in the semiclassical
approximation. 
In Eq.~\ceq{probfungauss}, which gives the second order corrections
to the classical action ${\cal A}_{cl}$, there are two Lagrange
multipliers, $\delta\lambda$ and $\lambda_{cl}$.
The most relevant condition is that imposed by
 $\delta\lambda$:
\beq{
\mathbf R'_{cl}\cdot\delta\mathbf R'=0
}{secondconstr}
Let us note that the above relation is at the origin of the asymmetry
in the generating functional
\ceq{finforgenfun}
between modes which are tangent or normal to the classical
background conformation $\mathbf R_{cl}$. 
Eq.~\ceq{secondconstr} is just the approximated version of the full
constraint \ceq{constbfr}. As a matter of fact, remembering the
splitting into classical solutions and statistical corrections of
Eqs.~\ceq{splitR} and \ceq{splitl}, we may rewrite Eq.~\ceq{constbfr}
as follows:
\beq{
(\mathbf R')^2_{cl}+2\mathbf R'_{cl}\cdot\delta \mathbf R'+
(\delta \mathbf R')^2-1=0
}{ccc}
Due to the fact that $(\mathbf R'_{cl})^2=1$ and neglecting the second
order term $(\delta \mathbf R')^2$, we obtain from Eq.~\ceq{ccc}:
\beq{
2\mathbf R'_{cl}\cdot\delta\mathbf R'=0
}{secondconstreq}
which coincides exactly with Eq.~\ceq{secondconstr}.
In the case of  solutions of type B there is an additional
constraint, which is associated to the nonzero constant Lagrange
multiplier $\lambda_{cl}$. This constraint
requires that 
the average over the time $t$ and
over the chain length $s$ of
the quadratic term  $(\delta \mathbf
R')^2$ is zero:
\beq{
\int_{t_i}^{t_f}\frac{dt}{t_f-t_i}dt\int_0^L\frac{ds}{L}(\delta
\mathbf R')^2=0
}{firstconstr}
In the solutions of type A the above condition is not present because
in that 
case $\lambda_{cl}=0$.
\section{The equilibrium limit of the GNLSM and its connection with
  the Rouse model }
%First of all, we study the equilibrium limit of the GNLSM. We use the
%formulation of the model given in
%Eqs.~(\ref{contparttd}--\ref{contacttd}). For simplicity, we put
%$t_i-0$.
First of all, we study the equilibrium limit of the GNLSM. We use the
formulation of the 
model given in Eqs.~(\ref{contparttd}--\ref{contacttd}). For simplicity,
we set $t_i=0$. 
Thus
\beq{
\Psi
=\int
{\cal D}\mathbf R(t,s)
e^{-\frac{M}{2L\kappa}
\int_0^{t_f}dt\int_0^Lds
\dot\mathbf R^2(t,s)
}
\delta\left(
\left|
\mathbf R^{\prime}(t,s)
\right|^2-1
\right)
}{newpart}
At this point, we rescale the time variable by putting
$\sigma=\frac{t}{t_f}$, so that 
the above equation becomes:
\beq{
\Psi
=\int{\cal D}\mathbf R(t_f\sigma,s)
e^{-\frac{M}{2L\kappa t_f}
\int_0^{1}d\sigma\int_0^Lds
\left(\frac{\partial\mathbf R}{\partial\sigma}\right)^2
}
\delta\left(
\left|
\mathbf R^{\prime}(t_f\sigma,s)
\right|^2-1
\right)
} {newpartbysigma}
In the equilibrium limit $t_f\rightarrow\infty$ we obtain:
\beq{
\Psi_{eq}
=\int{\cal D}\mathbf R_{eq}(s)
\delta\left(
\left|
\mathbf R_{eq}^{\prime}(s)
\right|^2-1
\right)
} {eqdist}
where  $\mathbf R_{eq}(s)=\mathbf R(\infty,s)$.
Eq.~\ceq{eqdist} is exactly what one should expect 
in  the case of the statistical mechanics of
a discrete chain 
 subjected to the constraints:
\beq{
\left|
\mathbf R_n - \mathbf R_{n-1}
\right|^2=a^2
} {wiezy}
The discrete probability function of the conformation of such a chain
is given by: 
\beq{
\Psi_{eq,disc}=
\prod_{n=1}^N
\int d\mathbf R_n\prod_{n=2}^N
\delta\left(
\frac{\left|\mathbf R_n - \mathbf R_{n-1}\right|^2}{a^2}
-1\right)
} {eqdistdisc}
In the continuous limit this becomes exactly the distribution of
Eq.~\ceq{eqdist}. This result
 is in agreement with the analogous probability function given in
\cite{EdwGoo}. 

While the purpose of this work is to provide a path integral
formulation of the dynamics 
of a freely jointed chain without having in mind concrete applications
to polymer physics, 
it is interesting to explore possible connections between the
GNLSM of 
Eqs.~(\ref{contparttd}--\ref{contacttd}) and the Rouse model. 
A direct attempt to put
the Rouse model in the path integral form
via the Martin-Siggia-Rose formalism leads to a probability distribution
for the Rouse chain 
which differs profoundly  from the GNLSM obtained in this work. Indeed,
let us
 start from the
Langevin equation:
\beq{
\zeta\frac{\partial\mathbf R}{\partial t}=\xi\frac{\partial^2\mathbf
  R}{\partial n^2} 
+\mathbf f
} {langevineq}
Here we have used instead of the arc-length $s$ the dimensionless
variable $n$ defined as follows: $s_0n=s$. Moreover,
$\mathbf f=\mathbf f(t,n)$ is a stochastic force with a Gaussian
distribution of width $\alpha$ given by:
$e^{-\int_{0}^{t_f}dt\int_0^{L\slash s_o}dn\frac{\mathbf
    f^2}{2\alpha}}$. $\zeta$ and $\xi$ are constant parameters which
will be specified later.
After the application of the
Martin-Siggia-Rose method,
one finds 
the Rouse
probability distribution:
\beq{
\Psi_{Rouse}
=\int
{\cal D}\mathbf R
e^{-\frac{1}{2\alpha}
\int_{0}^{t_f}dt\int_0^{L\slash s_0}dn
\left[
\zeta^2\left(\frac{\partial\mathbf R}{\partial t}\right)^2
+\xi^2\left(\frac{\partial^2\mathbf R}{\partial n^2}\right)^2
\right]
}
} {rousepd}
In principle, in the exponent of the above equation there should be  the
additional term 
\beq{
\displaystyle I=-\frac
{\zeta\xi}{\alpha}\int_{0}^{t_f}dt\int_0^{L\slash s_0}
dn\frac{\partial\mathbf R}{\partial t}\cdot\frac{\partial^2\mathbf
  R}{\partial n^2}}{addter}
However, due to the fact that $\frac{\partial \mathbf R}{\partial
  t}\cdot
\frac{\partial^2\mathbf R}{\partial n^2}=\frac{\partial}{\partial
  n}\left(
\frac{\partial \mathbf R}{\partial n}\cdot\frac{\partial\mathbf
  R}{\partial t}
\right)-\frac{\partial \mathbf R}{\partial n}\cdot \frac{\partial
  ^2\mathbf R}{\partial t\partial n}$ and remembering the identity
$\frac{\partial \mathbf R}{\partial n}\cdot \frac{\partial
  ^2\mathbf R}{\partial t\partial n}=\frac 12\frac{\partial}{\partial
  t}
\left[
\left(
\frac{\partial\mathbf R}{\partial n}
\right)^2
\right]
$, it is easy to realize that $I$ amounts  to  total derivative
terms, which can be neglected.

Coming back to Eq.~\ceq{rousepd}, we see that, while
the GNLSM is nonlinear and 
contains just second 
derivatives of the bond vector 
 $\mathbf R$, the Rouse model is linear and contains derivatives
of $\mathbf R$ 
up to the fourth order. As an upshot, while it is possible to investigate
the Rouse model 
by decomposing  $\mathbf R(t,s)$ into normal coordinates as
explained in 
\cite{doiedwards}, that kind of Fourier analysis cannot be easily
applied to the nonlinear 
GNLSM.
We show at this point that, indeed, the two models are quite different and
that it is not 
possible starting from one of them to recover the probability function
of the other and 
viceversa, because they correspond to regimes which do not overlap.
To this purpose, instead of the constraint $\mathbf R^{\prime 2}=1$ of
the GNLSM, we 
introduce the more general condition
\beq{
{\cal R}=
\left|
\frac{1}{s_0}\int_{-s_0}^{s_0}ds^{\prime} A(s^{\prime})
\left(
\mathbf R(t,s+s^{\prime})-\mathbf R(t,s)
\right)
\right|^2
-a^2=0
} {genconstr}
where $s_0$ is a new length scale such that
\beq{
a\ll s_0\ll L
} {socond}
and $A(s^{\prime})$ is a function of $s^{\prime}$ normalized in a such
a way that: 
\beq{
\frac{1}{s_0}\int_{-s_0}^{s_0}A(s^{\prime})ds^{\prime}=1
} {asnorm}
Let us note that $a$ is the smallest  length at our disposal:
Any segment of the chain 
of contour length shorter than $a$ may be regarded as rigid.
At this point, following the same strategy used in Section \ref{sec:three}, we
build the new 
distribution function
\beq{
\Psi_{int}
=\int{\cal D}\mathbf R
e^{-c\int_0^{t_f}dt\int_0^L ds  \dot\mathbf R^2}
\delta({\cal R})
} {xtl}
with $c=\frac{M}{4kT\tau L}$.
The index {\it int} means that the distribution probability
$\Psi_{int}$ describes 
a model which, as we will see, interpolates between the GNLSM and the
Rouse model. 

We remark that the insertion of the
$\delta$-function $\delta({\cal R})$ in
the path integral 
\ceq{xtl} has a double meaning.
On one side, it may be seen as a condition on the length of the
averaged vector: 
\beq{
\mathbf S = \frac{1}{s_0}
\int_{-s_0}^{s_0}ds^{\prime} A(s^{\prime})
\left(
\mathbf R(t,s+s^{\prime})-\mathbf R(t,s)
\right)
} {veces}
In the above equation
the distance between
points of the chain has been averaged over arc-segments of length $2s_0$. 
On the other side, the introduction of the $\delta$-function
$\delta({\cal R})$ may also 
be related to the internal forces among the beads composing the chain,
which appear 
due to the presence 
of constraints. For example, in the case of the GNLSM the presence of
these forces is 
evident in the formulation of Eq.~\ceq{pathlagmult}, in which the free
action is 
corrected by the 
addition of the interacting term $\lambda(\mathbf R^{\prime 2}-1)$. 

It is easy to realize that
 the GNLSM is a special case of the model described by Eq.~(\ref{xtl}). 
We have just to remember that in the GNLSM the motion of the chain is
observed at the 
smallest available 
scale of distances, i.~e. the segment length  $a$. 
Thus, we choose
the form of 
the function $A(s^{\prime})$ as follows:
\beq{
A(s^{\prime})=s_0\delta(s^{\prime}-a)
} {vershoran}
As a consequence, the constraint ~\ceq{genconstr} becomes
$\left|
\mathbf R(t,s+a)-\mathbf R(t,s)
\right|^2-a^2=0
$.
Dividing both members of the above equation by
$a^2$ and  supposing
that $a$ is very small, we get up to higher order terms in $a$
the relation:
\beq{\left|
\mathbf R^{\prime }
\right|^2-1= 0
} {bndfbgdh}
In the limit $a=0$, this is exactly the condition which has been
imposed to the chain in 
the GNLSM, see Eq.~\ceq{constbfr}. Using the property of the
$\delta$-function 
$\delta(a^2(|\mathbf R^{\prime}|^2 
-1))=\frac{1}{a^2}\delta(|\mathbf R^{\prime}|^2-1)$, it is possible to check
that also the probability 
distribution $\Psi_{int}$ becomes that of the GNLSM. 

To obtain the Rouse model from the interpolating probability
distribution $\Psi_{int}$, we
need first of all to decrease the resolution with which the segments
of the chain are observed. Accordingly, we require that the function
$A(s')$ appearing in the constraint \ceq{genconstr}  is constant over
the whole interval $[-s_0,s_0]$:
\beq{A(s')=\frac 12}{sdfsdf}
In this way, the finest details of the
chain are not taken into account, because the chain
conformations are averaged over the 
scale of distance $2s_0$, which is by hypothesis much larger than the
smallest available scale $a$. To pass to the Rouse model, we have also
to restrict 
ourselves to the 
long time-scale behavior of the chain.
This is achieved by assuming 
following~\cite{doiedwards}, Section 4.1, 
p.~93,
that, for long-time scales, $\mathbf R(t,s)$ varies slowly with $s$.
This hypothesis allows to stop the expansion of $\mathbf
R(t,s+s^{\prime})$ with 
respect to $s^{\prime}$ 
at the first few orders:
\beq{
\mathbf R(t,s+s^{\prime})=\mathbf R(t,s)+\mathbf R^{\prime}(t,s)s^{\prime}+
\frac{\mathbf R^{\prime\prime}(t,s)}{2}s^{\prime 2}+\ldots
} {Rexpansion}
Substituting the above truncated expansion in Eq.~\ceq{genconstr} and
performing the 
trivial integrations over $s^{\prime}$ we obtain the condition:
\beq{
\frac{|\mathbf R^{''}(t,s)|^2s_{0}^4}{6^2}-a^2=0
} {strcond}
Plugging in the above constraint in Eq.~\ceq{xtl}, we get the
following approximated 
expression of $\Psi_{int}$
\beq{
\Psi_{int}\sim
\int{\cal{D}}\mathbf R
e^{-c\int_0^{t_f}dt\int_0^Lds
\dot\mathbf R^2}
\delta(|\mathbf R^{\prime\prime }|^2s_0^4-a^2)
} {crst}
where the factor $6^2$ has been absorbed by a rescaling of the length
$s_0$. At this point 
we use the fact that, apart from an irrelevant infinite constant, the
functional 
$\delta$-function present in Eq.~\ceq{crst} may be simplified as follows:
\beq{
\delta(|\mathbf R^{''}|^2s_0^4-a^2)=\delta(|\mathbf R^{''}|s_0^2-a)
} {eqwwtp}
A proof of this identity, which is valid up to an irrelevant constant,
will be given in 
Appendix~B. Exploiting Eq.~\ceq{eqwwtp}, the probability distribution
$\Psi_{int}$ of 
Eq.~\ceq{crst} becomes:
\beq{
\Psi_{int}\sim\int{\cal{D}}\mathbf R
e^{-c\int_0^{t_f}dt\int_0^Lds
\dot\mathbf R^2
}\delta(\mathbf |\mathbf R^{''}|s_0^2-a)
} {atceq}
We may still simplify the above equation by applying 
the following slightly modified version of the gaussian approximation
of the $\delta$-function:
\beq{
\delta(|\mathbf R^{''}|s_0^2-a)\sim
\int{\cal{D}}\lambda
e^{
-i\int_0^{t_f}dt\int_0^Lds
\lambda\left(|\mathbf R^{''}|s_0^2-a\right)
}
e^{
-\int_0^{t_f}dt\int_0^Lds
\left(
\frac{\lambda^2}{2\nu}+i\frac{\beta}{\nu}\lambda
\right)
}
} {modfrep}
where we have supposed
 that the parameter $\nu$ is very large while the ratio
 $\beta\slash\nu$ is very small. Clearly, the usual Fourier 
representation  of the functional Dirac $\delta$-function is
recovered in the 
limit $\nu\rightarrow\infty$ and $\beta\slash\nu\longrightarrow
0$. 
Up to  now $\beta$ is an
arbitrary parameter.
We choose it in a such a way that
\beq{
\frac{\beta}{\nu}=a
} {gcalc}
This choice is compatible with our requirement for $\beta$,
since $a$ is the 
smallest  scale of lengths at our disposal, so that
$\beta\slash \nu$ is a very small quantity. Using Eq.~\ceq{gcalc} 
in order to eliminate $\nu$ 
from Eq.~\ceq{modfrep}, we obtain the relation:
\beq{
\delta(|\mathbf R^{''}|s_0^2-a)\sim
\int{\cal{D}}\lambda
e^{
-i\int_0^{t_f}dt\int_0^Lds
\lambda|\mathbf R^{''}|s_0^2
}
e^{-\int_0^{t_f}dt\int_0^Lds\frac{a}{2\beta}\lambda^2}
} {rclst}
After performing the gaussian integration over $\lambda$ in
Eq.~\ceq{rclst}, we 
obtain
\beq{
\delta(|\mathbf R^{''}|s_0^2-a)\sim
e^{-\int_0^{t_f}dt\int_0^Lds\frac{\beta}{2a}|\mathbf R^{''}|^2s_0^4}
} {fgfdhghfg}
We may now plug in the above expression of the $\delta$-function in
the distribution 
probability $\Psi_{int}$ of Eq.~\ceq{atceq}.
The result is:
\beq{
\Psi_{int}\sim
\int{\cal{D}}\mathbf R
e^{-\int_0^{t_f}dt\int_0^Lds
\left(
c\dot\mathbf R^2
+\frac{\beta}{2a}|\mathbf R^{''}|^2s_0^4
\right)
}
} {psiintrel}
This approximated probability distribution has the same structure of
that coming from the 
Rouse model given in Eq.~\ceq{rousepd}.

To make the comparison with the Rouse model more explicit, we perform
in Eq.~\ceq{psiintrel} the substitution $ns_0=s$:
\beq{
\Psi_{int}\sim \int {\cal D}\mathbf R
e^{-
\int_0^{t_f}dt
  \int_0^{L/s_0}dns_0
\left\{
c\dot{\mathbf R}^2+
\frac{\beta}{2a}\left|
\frac{\partial ^2\mathbf R}
{\partial n^2}
\right|^2
\right\}}
}{psint}
Let's now identify the coefficients appearing in Eq.~\ceq{psint} with
those of Eq.~\ceq{rousepd}.
We recall the fact that in the case of the Rouse model:
\beq{
\zeta=\frac 1\mu\qquad\qquad \xi = \frac{3
  kT}{s_0^2}\qquad\qquad\alpha=\frac{4kT}{\mu} 
}{eqparam}
On the other side, the parameter $c$ is  in the exponent of
Eq.~\ceq{psint} may be written as follows: $c=\frac{1}{4D}$.
It is now easy to verify that the probability function $\Psi_{Rouse}$
of Eq.~\ceq{rousepd} and that of Eq.~\ceq{psint} coincide if we make
the following choice for $\beta$: $\beta=\frac{9}{4}\frac{\mu a}{s^5_0}kT$.
\section{Conclusions}
This work may be considered as the ideal continuation of the seminal paper
of Edwards and Goodyear of Ref.~\cite{EdwGoo}, in which
the problem
of a chain subjected to the
constraints \ceq{constaconst} has been investigated using an approach
based on the
Langevin equation.
With respect to Ref.~\cite{EdwGoo}, our approach
based on the Fokker--Planck--Smoluchowski equation
provides
a path integral
and field theoretical formulation of  the dynamics of a freely jointed
chain in the continuous 
limit. The GNLSM obtained here makes possible the application of field
theoretical techniques to the
study of the fluctuations of a freely jointed chain. As an
example, we have derived 
in the semiclassical approximation
the probability function of the chain and the associated
generating functional. 
%The generating functional of the
%correlation functions of the vectors $\mathbf R(t,s)$ describing the
%spatial position of the points of the chain has been
%explicitly computed.  
 The approximation used in the computation of the
generating functional is
valid for instance in the cases of a cold isolated chain or of a
chain fluctuating in a very viscous medium. 

Most of the results obtained in this paper have been discussed in the
previous two Sections, so that we provide only a short summary of
them:
\begin{enumerate}
\item Derivation of the GNLSM, which provides a path integral
  formalism to the freely jointed chain.
\item Computation of the partition function and of the generating
  functional of the GNLSM in the semiclassical approximation.
\item The behavior of a chain at scales of length and time which are
very long
has been compared with the behavior at short scales of length and time 
in Section
  VII.
It is shown in this way that it is not possible to
 compare directly the Rouse model and models describing  a freely jointed
  chain like the GNLSM, because the regimes
and the assumptions of these two models do not overlap. In particular, the
Rouse model considers only the long-time behavior of the chain and long
scales of distances, while in the case of the freely jointed chain
 the short-time behavior is
taken into account and the chain is observed at a short scale of
 distance. 
\item A chain model which encompasses both the regimes of the Rouse
  model and of the GNLSM has been 
  proposed in Eq.~\ceq{xtl}.
\item The equilibrium limit of the GNLSM has been recovered. It gives
  the expected result in agreement with Ref.~\cite{EdwGoo}.
\item Last but not last, the
dynamics of a random chain has been investigated also from the
classical point of view. The equivalence of the
expressions of the classical Lagrangian of the chain computed starting
from cartesian and polar 
coordinates has been verified. 
\end{enumerate}

To conclude, we would like to mention
the problems 
which are still open and possible further developments of our work.
For simplicity, we have restricted our analysis to a two-dimensional
chain. However, its extension to any dimensions is not difficult.
Preliminary work in three dimensions
can be found in Ref.~\cite{FePaVi}. 
It turns out that more dimensions
allow the possibility of performing the continuous limit in different
ways, so that one could end up with a flexible chain or with a rigid
chain which privileges only certain directions along a fixed
axis. 
There should be also no problem  in switching on the interactions
among the beads composing the chain. To this purpose, one may use
the 
path integral methods applied to stochastic differential equations
explained in Ref.~\cite{zinn}. Only the inclusion of the hydrodynamic
interactions requires still some work. It is not simple to
provide for these interactions a lagrangian based formulation as that
developed here for a continuous chain. However, hydrodynamic interactions have
been already implemented in the path integral formalism in
Ref.~\cite{stepanow} using the Martin-Siggia-Rose formalism. Work is
in progress in order to extend the results of \cite{stepanow}  to the
freely jointed chain discussed in this work.
Another open question is how the chain behaves in
the short time regime when it is stretching under
the action of a force. This could be interesting
in the biophysics of DNA \cite{kroy}.
Finally, work is in progress in
order to linearize the GNLSM applying the same approach used in the
case of the standard nonlinear sigma model. In this way it would be
possible to study the GNLSM as the strong coupling limit of its linear
version with the help of the  techniques of~Refs.\cite{kleinertsc}.
\section{Acknowledgments} 
This work has been financed in part
by the Polish Ministry of Science, scientific project N202 156
31/2933.  
F. Ferrari gratefully acknowledges also the support of the action
COST~P12 financed by the European Union. 
The authors  would also like to thank
V. G. Rostiashvili for fruitful discussions and the anonymous referees
of Ref.~\cite{FePaVi} and of this work
for their precious remarks and suggestions.
Finally, F. Ferrari and J. Paturej are grateful 
for the nice
hospitality
to T. A. Vilgis and to
the Max Planck Institute for Polymer Research, Mainz, Germany.

\begin{appendix}
\section{Proof of Eq.~\ceq{funcid}}
In this Appendix we wish to prove Eq.~\ceq{funcid}. To this purpose,
we start from the path integral:
\beq{
I=\int {\cal D}x(t,s){\cal D}y(t,s)
f(x(t,s),y(t,s))
\delta\left(
(\partial_sx)^2+(\partial_sy)^2-1
\right)
}{idef}
Upon the transformation:
\beq{
x_s(t,s)=\partial_s x(t,s)\qquad\qquad y_s(t,s)=\partial_s y(t,s)
}{tranfoff}
we obtain:
\beq{
I=\int{\cal D}x_s(t,s){\cal D}y_s(t,s)\left(
\det{}^{-1}\partial_s
\right)^2f\left(
\int_0^sdux_u(t,u),\int_0^sduy_u(t,u)
\right)\delta\left(
x_s^2+y_s^2-1
\right)
}{idel}
where we have made use of the fact that:
\beq{
\det\left(\frac{\delta x}{\delta x_s}\right)
\det\left(\frac{\delta y}{\delta y_s}\right)=\left(
\det\partial_s^{-1}
\right)^2=\left(
\det{}^{-1}\partial_s
\right)^2
}{ideaaa}
Now it is possible to eliminate the variable $y_s$ by performing
 in $I$ the substitution:
\beq{
\chi=x_s^2+y^2_s-1
}{substfd}
As a consequence, the path integration over $y_s$ appearing in
\ceq{idel} may be replaced by a path integration over the new variable
$\chi$:
\begin{eqnarray}
\int
{\cal D}y_s\delta\left(
x_s^2+y_s^2-1
\right)f\left(\int_0^sx_udu,\int_0^sy_udu\right)=&&\nonumber\\
\left.\int_{\chi\ge x_s^2-1}{\cal D}\chi\delta(\chi) \det\left|
\frac{\delta y_s}{\delta \chi}\right|f\left(
\int_0^sx_udu,\int_0^sy_udu
\right)\right|_{y_s=\pm\sqrt{1+\chi-x_s^2}}\label{idetwo}
\end{eqnarray}
In the above equation the determinant $\det\left|
\frac{\delta y_s}{\delta \chi}\right|$ is the functional determinant
giving the Jacobian of the transformation \ceq{substfd}, so that:
\beq{\det\left|
\frac{\delta y_s}{\delta \chi}\right|=\det\left(
\frac{1}{2\sqrt{1+\chi-x_s^2}}
\right)}{jactrans}
Applying Eqs.~\ceq{ideaaa} and \ceq{idetwo} in \ceq{idef} we get:
\beq{
I=\int_{x_s^2-1\le0}{\cal
  D}x_s\det{}^{-1}\left|\partial_s\right|^2\det{}^{-1} \left(
2\sqrt{1-x_s^2}
\right)f\left(
\int_0^sx_u(t,u)du,\pm\int_0^s \sqrt{1-x_u^2(t,u)}
\right)
}{lastbeforesincos}
Finally, we perform in Eq.~\ceq{lastbeforesincos} the substitution
$x_s=\cos\varphi$. In this way $I$ may be rewritten as a path integral
over $\varphi$:
\beq{
I=\int{\cal D}\varphi\det\left|\frac{\delta x_s}{\delta\varphi}\right|
\det{}^{-1}\left|\partial_s^2\right|\det{}^{-1}\left(
2\sin\varphi
\right)f\left(
\int_0^s\cos\varphi(t,u)du,\int_0^s\sin\varphi(t,u)du
\right)
}{resvar}
Noting that $\det\left|\frac{\delta
  x_s}{\delta\varphi}\right|=-\sin\varphi$ we obtain, apart from an
irrelevant constant ${\cal
  N}=\det{}^{-1}\left|\partial_s^2\right|\det{}^{-1} 2$, the final result:
\beq{I={\cal N}\int{\cal D}\varphi
f\left(
\int_0^s\cos\varphi(t,u)du,\int_0^s\sin\varphi(t,u)du
\right)
}{finres}
\section{A functional identity}
To prove the identity \ceq{eqwwtp}, we write the functional
$\delta-$function in the right hand side of Eq.~\ceq{eqwwtp} with the
help of its Fourier representation:
\beq{
\delta\left(
|\mathbf R''|^2s_0^4-a^2
\right)=\int{\cal D}\tilde\lambda
e^{-i\int_0^{t_f}dt\int_0^Lds\tilde\lambda
(|\mathbf R''|^2s_0^4-a^2)}
}{frep}
Since $|\mathbf R''|^2s_0^4-a^2= (|\mathbf R''|s_0^2-a)(|\mathbf
R''|s_0^2+a )$, Eq.~\ceq{frep} becomes:
\beq{
\delta\left(
|\mathbf R''|^2s_0^4-a^2
\right)=\int{\cal D}\tilde\lambda
e^{-i\int_0^{t_f}dt\int_0^Lds\tilde\lambda
(|\mathbf R''|s_0^2-a)(|\mathbf
R''|s_0^2+a ) }
}{freppol}
Due to the fact that $|\mathbf
R''|s_0^2+a $ is always different from zero, one may perform the
change of variables:
\beq{
\tilde\Lambda =\tilde\lambda( |\mathbf
R''|s_0^2+a  )
}{substllt}
Applying the substitution \ceq{substllt} in \ceq{freppol} we obtain:
\beq{
\delta\left(
|\mathbf R''|^2s_0^4-a^2
\right)={\det}^{-1}(2a)\delta(R''|s_0^2-a)
}{dgfhsdgfhg}
which coincides with Eq.~\ceq{eqwwtp} up to the irrelevant constant
$\det{}^{-1}(2a)$.
\end{appendix}

\end{document}